\documentclass[12pt,draftclsnofoot,onecolumn]{IEEEtran}

\usepackage{amsmath}
\usepackage{amsthm}
\usepackage{amsfonts}
\usepackage{amssymb}
\usepackage{makeidx}
\usepackage{cite}
\usepackage{graphicx}
\usepackage{mathtools}
\usepackage{cases}
\usepackage{xcolor}

\usepackage{hyperref}
\usepackage{bbm}
\usepackage[normalem]{ulem}
\usepackage{enumerate}   
\usepackage{bigints}
\usepackage{epstopdf}
\usepackage{braket}
\usepackage{graphicx}

\newtheorem{theorem}{Theorem}

\newtheorem{lemma}{Lemma}

\newtheorem{example}{Example}

\newtheorem{remark}{Remark}

\DeclareMathOperator*{\argmax}{arg\,max}

\DeclareMathOperator{\cL}{\mathcal{L}}

\DeclareMathOperator{\cT}{\mathcal{T}}

\DeclareMathOperator{\SIR}{\textrm{SIR}}
\DeclareMathOperator{\bR}{\mathbb{R}}

\DeclareMathOperator{\bP}{\mathbf{P}}
\DeclareMathOperator{\ind}{\mathbbm{1}}
\DeclareMathOperator{\bE}{\mathbf{E}}

\newcommand*\diff{\mathop{}\!\mathrm{d}}

\newcommand*\nnb{\nonumber}
\newcommand*\nnnl{\nonumber\\}

\newcommand{\ea}{\stackrel{(\text{a})}{=}}
\newcommand{\eb}{\stackrel{(\text{b})}{=}}
\newcommand{\ec}{\stackrel{(\text{c})}{=}}
\newcommand{\ed}{\stackrel{(\text{d})}{=}}
\newcommand{\ee}{\stackrel{(\text{e})}{=}}
\newcommand{\ef}{\stackrel{(\text{f})}{=}}
\newcommand{\eg}{\stackrel{(\text{g})}{=}}
\newcommand{\eh}{\stackrel{(\text{h})}{=}}

\title{Spatial and Temporal Analysis\\of Direct Communications\\from Static Devices to Mobile Vehicles}
\author{Chang-Sik~Choi~and~Fran{\c{c}}ois~Baccelli
	\thanks{Chang-Sik~Choi and Fran{\c{c}}ois~Baccelli are with the Wireless Networking and Communication Group, Department of Electrical and Computer Engineering, The University of Texas at Austin, TX, USA (email: chang-sik.choi@utexas.edu,  baccelli@math.utexas.edu)}
	\thanks{A short version of the manuscript is in Proc. IEEE Globecom 2018 \cite{ccs}.}
	\thanks{Last revised: \today}}
\begin{document}
	\maketitle
	\begin{abstract}
 This paper proposes a framework to analyze an emerging wireless architecture where vehicles collect data from devices. Using stochastic geometry, the devices are modeled by a planar Poisson point process. Independently, roads and vehicles are modeled by a Poisson line process and a Cox point process, respectively. For any given time, a vehicle is assumed to communicate with a roadside device in a disk of radius $ \nu $ centered at the vehicle, which is referred to as the coverage disk. We study the proposed network by analyzing its short-term and long-term behaviors based on its space and time performance metrics, respectively. As short-term analysis, we explicitly derive the signal-to-interference ratio distribution of the typical vehicle and the area spectral efficiency of the proposed network. As long-term analysis, we derive the area fraction of the coverage disks and then compute the latency of the network by deriving the distribution of the minimum waiting time of a typical device to be covered by a disk. Leveraging these properties, we analyze various trade-off relationships and optimize the network utility. We further investigate these trade-offs using comparison with existing cellular networks.
	\end{abstract}
\section{Introduction}
\subsection{Motivation and Background}
This paper studies an emerging wireless architecture where devices collect data and passing-by vehicles harvest their data. The idea of using vehicles as key network components---like base stations or access points---is widely investigated from both industry \cite{forstall2013mobile,saad2014vehicle,talluri2018enhanced} and academia \cite{Hull:2006:CDM:1182807.1182821,jain2006exploiting,Xing:2008:RDA:1374618.1374650,Choi:2018:DLM:3209582.3209590}. Concrete examples of such networks range from ad hoc networking (where  public transit vehicles provide  large-scale Internet connectivity for pedestrians \cite{Veniam}), to vehicular-to-all---or equivalently device-to-device---networks (where pedestrians' mobile devices send safety information to nearby vehicles \cite{sugimoto2008prototype,anaya2014vehicle}), and to Internet-of-Things (IoT) networks (in which roadside sensors opportunistically forward their data toward nearby vehicles \cite{jain2006exploiting,Choi:2018:DLM:3209582.3209590}). These examples share the basic idea that the devices are distributed in space and vehicles directly collect the data from neighboring devices. 
\par  
This paper proposes a network model based on direct communications from static data devices to dynamic vehicles and then analyzes its performance.  Specifically, it proposes a stochastic geometry model \cite{baccelli2010stochastic,chiu2013stochastic}. The analysis presented in this paper sheds light on the performance of the proposed network and more generally provides a framework to quantify the potential of network architectures leveraging vehicles.

\subsection{Related Work}
The proposed network architecture is an example of random mobile ad hoc network or device-to-device network in the sense that it can expand the limited coverage of infrastructure or enable high-speed and low-distance communication between devices without infrastructure \cite{hartenstein2008tutorial,huang2009spectrum,doppler2009device,golrezaei2013femtocaching,feng2013device,andreev2014cellular}. The performance of these networks has been studied extensively, with some studies using stochastic geometry to model the random locations of network components \cite{weber2005transmission,baccelli2006aloha,andrews2010primer,lin2014overview}. For instance, the homogeneous planar Poisson point process has been widely used for its analytical tractability \cite{haenggi2009stochastic,andrews2010primer}.  Specifically, under the Palm distribution of the Poisson point process, the distribution of the signal-to-interference-plus-noise ratio (SINR) of a typical user and the network area spectral efficiency were derived in  \cite{baccelli2006aloha,baccelli2009stochastic,ganti2009spatial}.

However, modeling the locations of vehicles as a planar Poisson point process is inaccurate since almost surely no more than two points can be found on a line in the planar Poisson point process \cite{moller2012lectures}, and yet the locations of vehicles exhibit a linear pattern when they are on the same straight road. In order to address the location dependencies, a Poisson-line Cox model was proposed in \cite{baccelli1997stochastic}, where roads and vehicles are conditionally generated in the Euclidean plane. More recently, this model was further studied  in \cite{morlot2012population,choi2017analytical,chetlur2017coverage} to derive the signal-to-interference ratio (SIR) distribution of various links between vehicles and mobiles on the plane. These papers analyzed the typical network performance by considering an instantaneous snapshot of the network geometry, under the Palm distribution of the vehicle point process. This paper uses the same approach to characterize short-term performance properties such as the distribution of the SIR and the area spectral efficiency.

\par On the other hand, since vehicles are assumed to cover a wide area as they move on roads, it is essential to analyze the network behavior over time. This paper uses the theory of random closed sets \cite{molchanov2005theory,chiu2013stochastic} to derive 
the area fractions of the coverage disks and of the progress of coverage over time, respectively. In addition, as in the literature on delay-tolerant networks \cite{grossglauser2002mobility,zorzi2003geographic,fall2008dtn,skordylis2008delay,pereira2012delay} or on random networks with data mules \cite{shah2003data,spyropoulos2005spray,zhao2005controlling}, in the proposed network users might incur additional delay for link association when the density of vehicle is small or the speed of vehicle is slow. To quantify this association delay, this paper investigates the network latency by deriving the distribution of the shortest time for a typical roadside device to be covered by any vehicle, or equivalently any disk.

\subsection{Contributions}
\textbf{Modeling of the proposed network}: The paper considers a generic architecture where data devices communicate with vehicles on roads.  The devices are modeled by a planar Poisson point process with high density. Independent of the device point process, a Poisson line process and conditional linear Poisson point processes on each line model the road network and the vehicles on the roads, respectively. At any given time, a vehicle communicates with at most one device in the coverage disk centered at the vehicle. Vehicles are assumed to move along the lines of the Poisson line process at a constant speed, and thus vehicles collect the data from various devices as they move.

\textbf{Performance analysis for short-term performance behavior}: The short-term performance is analyzed by considering a snapshot of the proposed stochastic geometry model. To be specific, using the Palm distribution of the vehicle point process and assuming rich scattering and a general power-law path loss, we obtain  integral formulas for the interference distribution at a typical vehicle and the SIR coverage probability of the typical vehicle. By deriving the ergodic throughput of the typical link, we obtain the expression for the area spectral efficiency.

\textbf{Performance analysis for long-term performance behavior}: The long-term performance is analyzed by studying the evolution of the coverage disks over time. Specifically, the evolution of the coverage disks is characterized as the Minkowski sum \cite{molchanov2005theory} of the trajectories of vehicles and the coverage disk. Using the stationarity of the associated random closed sets and the capacity functional formula, we explicitly derive the area fraction of the coverage disks and its evaluation over time. Conditionally on the fact that the distance from a typical device to any road is less than the radius of the coverage disk, the network latency is defined as the link association delay, namely, the amount of time that a typical device has to wait in order to be covered by a disk. We derive its distribution and mean value.  
 
\textbf{Trade-off relationship, optimization, and comparison}: We obtain various trade-off relationships between short-term and long-term properties. For instance, when the coverage disk radius increases (resp. decreases), the long-term performance metrics, e.g., the area fraction and the latency improve (resp. worsen) while the short-term performance metrics, e.g., the coverage and rate worsen (resp. improve). We find a similar trade-off with respect to the linear density of vehicles. To shed light on the potential of the proposed architecture, we consider a utility function that incorporates these key metrics and we optimize it with respect to the coverage disk radius. To further evaluate the proposed network, we compare it with cellular architecture based on hexagonal grids.


	\section{System model}\label{S:2}
	This section provides the spatial model for the devices, vehicles, and coverage disks. Then, the channel model and performance metrics are discussed.
	\subsection{Spatial Model}\label{S:2-A}
	
First, we assume that the devices are distributed according to an independent planar Poisson point process with density $ \lambda $ in the Euclidean plane $ \bR^2 $. The devices are full-buffered and they communicate with vehicles. We assume $ \lambda $ to be high.
	
We model the road network using a stationary Poisson line process $ \Phi_l $ with intensity $ \lambda_l $, independent of the device point process. The Poisson line process  is generated as follows: a Poisson point process with intensity $ \lambda_l $ is generated on the cylinder set $ \mathbf{C}:\bR\times (0,\pi)$.  Then, each point of the Poisson point process, say $ (r,\theta), $ gives birth to  a line on the Euclidean plane $ \bR^2 $ where $ r $ describes the distance from the origin to the line and $ \theta $ is the argument of the normal vector to the line. 
\par Conditionally on the line process $ \Phi_l $, the locations of vehicles on each road are modeled by an independent one-dimensional Poisson point process with intensity $ \mu $ on each line. The collection of vehicles on the line process hence forms a Poisson line Cox point process $ \Phi $ \cite{choi2018poisson}. The spatial density of the Poisson line Cox point process, or equivalently its intensity, is equal to $ \lambda_l\mu $\cite[Lemma 2]{choi2018poisson}, the product the road density $ \lambda $ and the linear density $ \mu. $
\par In order to describe the motion of vehicles, we consider a simple dynamic. Each vehicle is assumed to move at a constant speed $ v $ along its line. The direction of motion is randomly determined by an independent coin tossing at each vehicle. Once it is determined, each vehicle maintains its initial direction and speed. Let $ \Phi(t) $ denote the locations of vehicles at time $ t $.

\begin{figure}
	\centering
	\includegraphics[width=.7\linewidth]{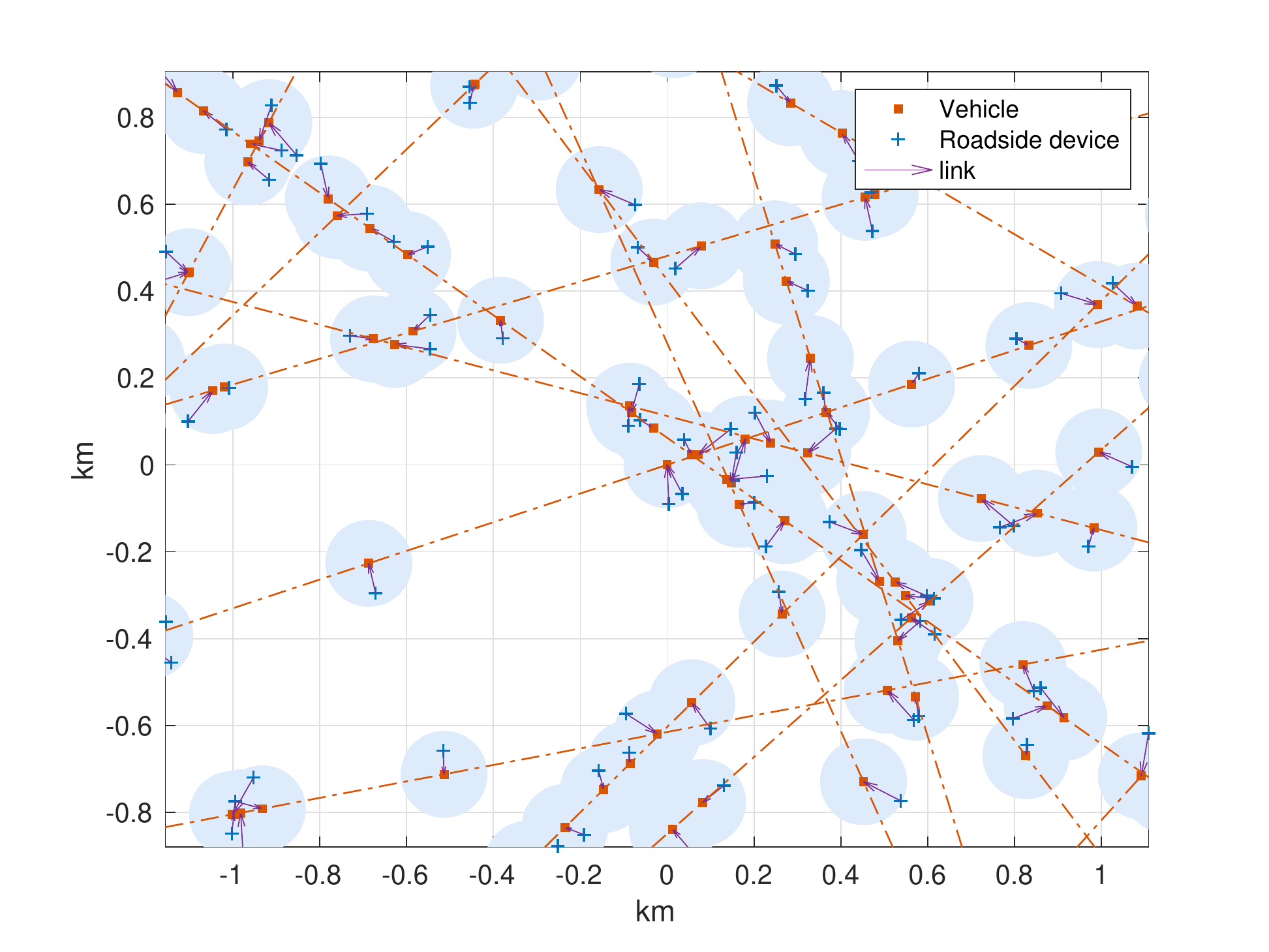}
	\caption{Illustration of the proposed network architecture where roads are distributed according to a Poisson line process.}
	\label{fig:proposednetwork}
\end{figure}

\begin{figure}
	\centering
	\includegraphics[width=.7\linewidth]{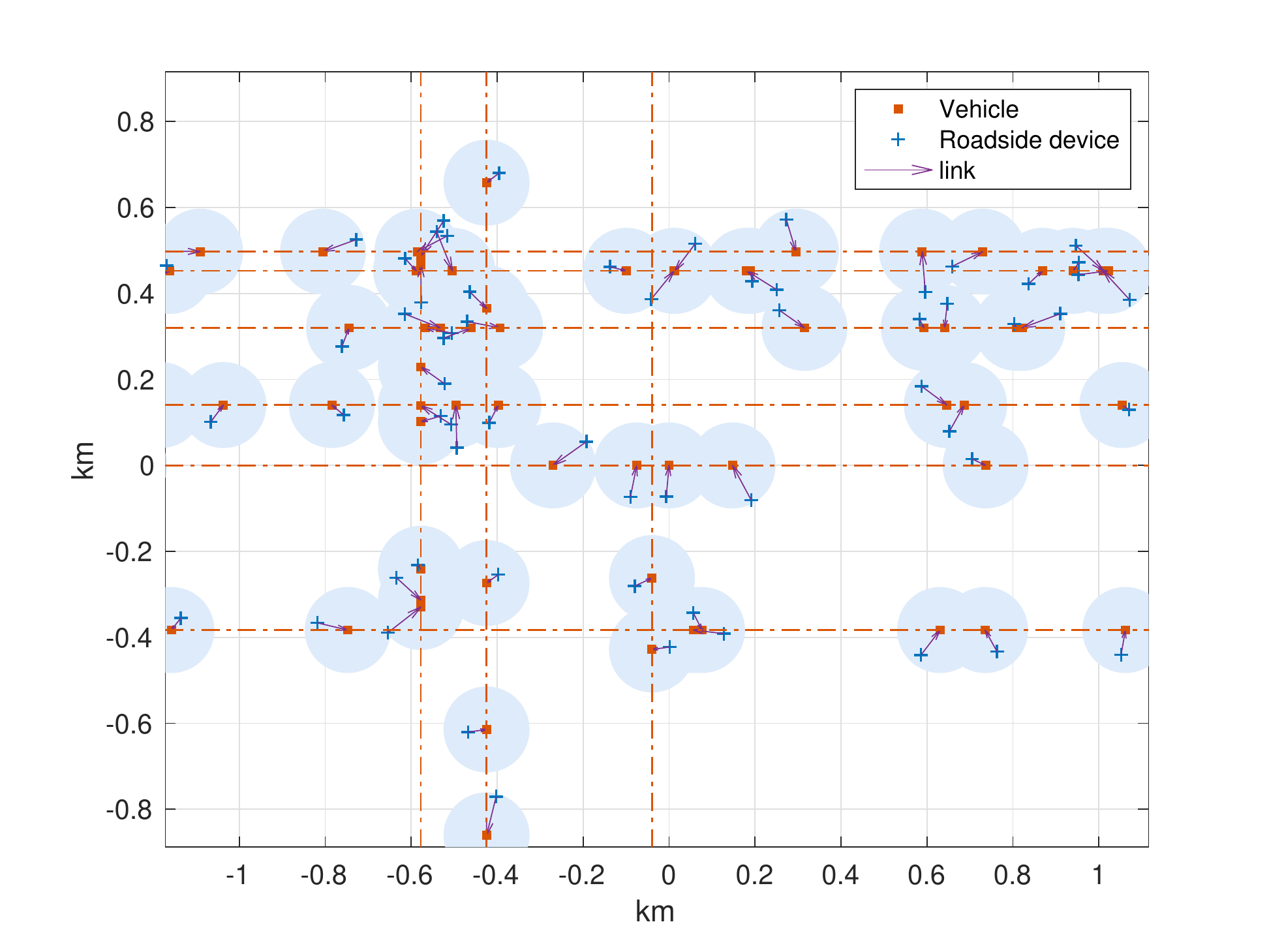}
	\caption{Illustration of the proposed network architecture with the Manhattan Poisson line process.}
	\label{fig:proposednetwork_manhattan}
\end{figure}

\subsection{Coverage Disks}
\par Each vehicle has a disk of radius $ \nu $ centered at its location, which is referred to as the coverage disk. The devices inside the disk are \emph{geometrically} covered by the vehicle. Since vehicles change their locations over time, so do the coverage disks. Let $ \cup_{X\in\Phi(t)} B_{X}(\nu) $ denote the collection of the coverage disks where $ B_{X}(\nu) $ denotes the disk of radius $ \nu $ centered on the vehicle located at $ X $. Time is slotted and we assume that at any given time, one device inside each disk is selected to transmit according to some time division multiple access mechanism.

\par Since we assume that the device density is high, the probability that each disk has no device is almost zero:  $ 1-\exp(\pi\lambda \nu^2)\approxeq 0 $. Given that there exists at least one data device per coverage disk, the location of each selected device is uniformly distributed in the disk due to the Poisson property \cite{baccelli2009stochastic}. From this point in the paper, we use the term \emph{roadside devices} to refer to the devices selected in the coverage disks.
\par The point process for the roadside devices at any given time is given by 
\begin{equation*}
\Xi=\sum_{X_i\in\Phi}\Xi_{X_i}=\sum_{X_i\in\Phi,U_i\in \text{Uniform}(B(\nu))}\delta_{X_i+U_i},   
\end{equation*}
where $ \delta_{x} $ denotes the Dirac-delta function of mass one at location $ X $ and $ \{U_i\} $ is an i.i.d. sequence of vectors, uniformly distributed in the disk $ B_0(\nu) $. 

Fig. \ref{fig:proposednetwork} illustrates the vehicle point process, the coverage disks, and the roadside devices. We consider $ \lambda_l=3/\text{km}, $ $ \mu=3/\text{km}, $ and $ \nu=100 \text{m}. $ In a similar way, Fig. \ref{fig:proposednetwork_manhattan}  illustrates the proposed network where roads are now modeled by the Manhattan Poisson line process \cite{chiu2013stochastic} (roads parallel to the $ x $- and $ y $-axes are produced by restricting $ \theta $-values to $ 0 $ or $ \pi/2 $ \cite[Fig. 1]{choi2018poisson}). In both figures, a typical vehicle is considered at the origin.

\subsection{Propagation Model}
To characterize the received signal power at the vehicles, we consider a classical power law path loss model with Rayleigh fading. The received power at distance $ d $ is $ pHd^{-\alpha}, $ where $ p $ is the device transmit power, $ H $ is the power of Rayleigh fade that follows an exponential distribution with mean one, and $ \alpha $ is the path loss exponent ($ \alpha>2 $).

\begin{table}
	\caption{Network parameters and performance metrics}\label{T:1st}
	\centering
	\begin{tabular}{|l|l|}
		\hline
		Parameters & Notations $ \sim $ Distribution\\
		\hline 
		Roads &  $ \Phi_l \sim \text{Poisson line with density }\lambda_l$\\ 
		\hline 
		Vehicles on each line & $ \phi \sim $  Poisson process with density $ \mu $\\
		\hline 
		Vehicles at time $ t $	& $ \Phi(t)\sim \text{Cox point process}$ \\ 
		\hline
		Ball (disk) of radius $ r $ centered at $ x $ & $ B_{x}(r) $\\\hline
		Roadside device around vehicle $ X $ & $ U_j\sim \text{Unif}(B_X(\nu)) $ \\ 
		\hline 
		Union of coverage disks at time $ t $& $ S(t)=\bigcup\limits_{X\in\Phi(t)} B_{X}(\nu)$ \\
		\hline 
		Cumulative covered area  up to time $ t $& $ \bar{S}{(t)} = \bigcup\limits_{0\leq \nu \leq t}\bigcup\limits_{X\in\Phi(\nu)} B_{X}(\nu)$  \\ 
		\hline
		Laplace transform of the interference & $ \cL_{I}(s) $ \\
		
		\hline 
		Coverage probability with threshold $ \tau $& $ p_c(\tau) $\\ 

		\hline
		
	\end{tabular} 
\end{table}

\subsection{Performance Metrics}
\subsubsection{Space Domain} We focus on the typical link, namely a link from a roadside device to the typical vehicle at the origin. For a discussion on the Palm distribution of the proposed Poisson line Cox point process, see e.g., \cite{morlot2012population,choi2017analytical}. We consider the coverage probability of a typical vehicle and then the area spectral efficiency of the network.  

	\textbf{Coverage probability}: This is defined by the probability that the SIR of the typical vehicle to be larger than some threshold $ \tau  $, i.e., the complementary cumulative distribution function of the SIR random variable. The coverage probability is defined by 
	\begin{align}\label{eq:2}
	p_c(\tau)=\bP_{\Phi}^{0}(\SIR\geq \tau),
	\end{align}
	where $ \bP_{\Phi}^{0} $ denotes the Palm distribution with respect to the vehicle point process $ \Phi $. The coverage probability can be interpreted as the spatial fraction of links whose SIRs are larger than the threshold.

	\textbf{Area spectral efficiency}: Using the Shannon formula, the area spectral efficiency is  
	\begin{align}\label{eq:4}
	\text{ASE}=\lambda_l\mu \bE_{\Phi}^{0}\left[ \log_2(1+\SIR)\right],
	\end{align}
where the expectation is with respect to the Palm distribution of $ \Phi. $

\subsubsection{Time Domain}
We consider the mean area fraction and the shortest waiting time. Both metrics are related to the behavior of the coverage disks. Figs. \ref{fig:proposednetwork} and \ref{fig:proposednetwork_manhattan} illustrate the coverage region as a union of the coverage disks. The mean area fraction of the stationary random set $ A $ is defined as the probability that the origin lies in $ A $\cite{daley2008introduction}. From this point of the paper, we will refer to the mean area fraction as the area fraction, for simplicity. The area fraction of the coverage disks is also given by  
\begin{equation*}
\text{AF}(S(t))=\lim_{r\to\infty}\frac{\bE[\ell_2(S(t)\cap B_0(r))]}{\ell_2(B_0(r))},
\end{equation*}
where $ \ell_2(A) $ denotes the area of set $ A $ and $ B_0(r) $ is the disk of radius $ r $ centered at the origin. The area fraction is always between $ 0 $ and $ 1. $ 

	\textbf{Area fraction of the cumulative coverage disks}: The union of all coverage disks between time $ 0 $ and time $ t $ is given by 
	\begin{equation}
	\bar{S}{(t)}=\bigcup_{0\leq \nu \leq t}S(\nu)=\bigcup_{0\leq \nu\leq t}\left(\bigcup_{X\in\Phi(\nu)}B_{X}(\nu)\right).
	\end{equation}  
	Therefore, the area fraction of the set $ \bar{S}(t) $ is given by
	\begin{equation}\label{}
	\text{AF}({\bar{S}}(t))=\lim_{r\to\infty}\frac{\bE[\ell_2(\bar{S}{(t)}\cap B(r))]}{\ell_2( B(r))}.
	\end{equation}

	 \textbf{Network latency}: The network latency is defined on the event that the typical device is contained in set $ \bar{S}(t) $ at time infinity. We define the waiting time of the typical device to be covered by a coverage disk as follows:
	\begin{equation}\label{6-1}
	W=\left.\inf_{\tau>0}\left\{\tau \text{ such that } \bar{S}(\tau)\cap 0 \neq \emptyset  \right| 0 \in \bar{S}(\infty)\right\}.
	\end{equation}
	The network latency is defined as the mean waiting time. 

\section{Performance Analysis in the Space Domain}\label{S:3}
This section focuses on the space domain using the instantaneous layout of the network geometry. The space domain metrics capture the short-term behavior of the proposed architecture.
\subsection{Interference at Vehicle}
Roadside devices directly communicate with vehicles and they are uniformly located in the coverage disks centered at vehicles. Hence, we focus on the distribution of the interference power measured at vehicles. Specifically, we derive the interference under the Palm distribution,  considering the typical vehicle at the origin.

\begin{lemma}\label{L:1}
The Laplace transform of the interference at the typical vehicle is given by
	\begin{align}
		\cL_{I}(s)=&\exp\left(-\lambda_l\int_{\bR}1-e^{-\frac{\mu}{\pi \nu^2}\int_{\bR}\int_{B_0(\nu)}\frac{sp{((r+u)^2+(t+v)^2)}^{-\alpha/2}}{1+sp{((r+u)^2+(t+v)^2)}^{-\alpha/2}}\diff u\diff v\diff t}\diff r\right)\nnb\\
		&\exp\left(-{\mu}\int_{\bR}\int_{B_{0}(\nu)}\frac{sp{((t+u)^2+v^2)}^{-\alpha/2}}{1+sp{((t+u)^2+v^2)}^{-\alpha/2}}\diff u\diff v \diff t\right)\label{6},
	\end{align}
	where $ \lambda_l $ is the road density and $ \mu $ is the linear density of vehicle on each road. 
\end{lemma}

\begin{IEEEproof}
We consider a typical vehicle located at the origin. Consequently, a typical line at the origin exists \cite{choi2018poisson}. We denote by $ \phi(0) $ the Poisson point process with intensity $ \mu $ on the typical line. Then, under the Palm distribution of the vehicle point process, the interference can be decomposed as follows:
		\begin{align*}
		I		 =\underbrace{\sum_{X_i\in\Phi}p H {\|X_i+U_i\|}^{-\alpha}}_{I_1}+\underbrace{\sum_{X_i\in\phi(0)}p H {\|X_i+U_i\|}^{-\alpha}}_{I_2},
	\end{align*}
	where $ I_2 $ denotes the interference from all roadside devices activated by vehicles on the line at the origin i.e., the typical line and $ I_1 $ accounts for the interference from all roadside devices activated by vehicles on the rest of the lines. Due to Slivnyak's theorem applied to the Poisson point process on the cylinder set, the points on the typical line---considered under the Palm distribution of the vehicle point process---and the other points on the rest of the lines are independent \cite{choi2018poisson}. 
	\par Consequently, the random variables $ I_1 $ and $ I_2 $ are independent and the Laplace transform of the interference is given by
\begin{align}
\bE\left[\exp(-sI)\right]&=\underbrace{\bE[\exp(-sI_1)]}_{\cL_{I_1}(s)}\underbrace{\bE[\exp(-sI_2)]}_{\cL_{I_2}(s)}.
\end{align}
	
\par To begin with, the Laplace transform of random variable $ I_1 $ is given by 
\begin{align}
	\cL_{I_1}(s)&=\bE_{\Phi}\left[\prod_{X_i\in\Phi}\bE_{{U}}\left[\left.\bE_H\left[e^{-spH\|X_i+U\|^{-\alpha}}\right|\Phi ,U\right]\right]\right]\nnb\\
	&\ea\bE_{\Phi}\left[\prod_{X_i\in\Phi}\bE_{U}\left[\left.\frac{1}{1+sp \|X_i+U\|^{-\alpha}}\right|\Phi\right]\right]\nnb\\
	&=\bE_{\Phi}\left[\prod_{X_i\in\Phi}\frac{1}{\ell_2(B_0(\nu))}\int_{B_0(\nu)}\frac{1}{1+sp{\|X_i+\mathbf{u}\|}^{-\alpha}}\diff \mathbf{u}\right].\label{le2}
\end{align} 
In order to obtain (a), we use that the locations of roadside devices are given by $ Y_i=X_i+U_i $ where $ \{U_i\} $ denote vectors independent and identically distributed (i.i.d.) in disk $ B_0(r) $.

\par Then, the locations of all points of $ \Phi $ can be denoted by the summation of vectors $ X_i=r_j\vec{\rho_j}+t_k\vec{\kappa_k} $ where $ \vec{\rho_j} $ is the unit vector normal to the line of $ X_i $ and $ \vec{\kappa_j} $ is a unit vector of the line of $ X_i $.  Here, $ r_j $ and $ t_k $ correspond to the distances from the origin to the $ X_i $,  with respect to the vector $ \rho_j $ and $ \kappa_j $, respectively. 
See \cite{choi2018poisson} for a measurable enumeration of points of the Poisson line Cox point process. Here, 

Then, the random vector $ \mathbf{u} $ of Eq. \eqref{le2} can be written as $ u\vec{\rho_j}+ v \vec{\kappa_j} $ using the above orthonormal vectors. As a result, the Laplace transform is given by 
\begin{align}
	&\bE_{\Phi}\left[\left.\prod_{X_i\in\Phi}{}\int_{B_0(\nu)}\frac{1}{\pi \nu^2}\frac{1}{1+sp{\|X_i+\mathbf{u}\|}^{-\alpha}}\diff \mathbf{u}\right.\right]\nnb\\
	&=\bE_{\Phi_l}\left[\prod_{r_j}\bE\left[\left.\prod_{t_k\in\phi(r_j)}\int_{B_{}(\nu)}\frac{1}{\pi \nu^2}\frac{1}{1+sp{\|(r_j+u)\vec{\rho_j}+(t_k+v)\vec{\kappa_j}\|}^{-{\alpha}}}\diff u\diff v\right|\Phi_l\right]\right]\nnb\\
	&\eb\bE_{\Phi_l}\left[\prod_{r_j}\bE\left[\left.\prod_{t_k\in\phi(r_j)}\int_{B_{}(\nu)}\frac{1}{\pi \nu^2}\frac{1}{1+sp{((r_j+u)^2+(t_k+v)^2)}^{-\frac{\alpha}{2}}}\diff u\diff v \right|\Phi_l\right]\right]\nnb\\
	&\ec\bE_{\Phi_l}\left[\left.\prod_{r_j}\exp\left(-\frac{\mu}{\pi \nu^2}\int_{\bR}\int_{B_0(\nu)}\frac{sp{((r_j+u)^2+(t+v)^2)}^{-\frac{\alpha}{2}}}{1+sp{((r_j+u)^2+(t+v)^2)}^{-\frac{\alpha}{2}}}\diff u\diff v\diff t\right)\right.\right]\nnb.
			\end{align}
		In order to derive (b), we use the fact that $ \vec{\rho_j} $ and $ \vec{\kappa_j} $ are orthonormal vectors. To have (c), we use the probability generating functional of the Poisson point process $ \phi(r_j) $ with linear intensity $ \mu. $ In the remainder of this paper, we use the single integral notation $ \int_{B_{0}(\nu)} \cdot \diff u \diff v $ to represent the double integral $ \int\int_{B_0(\nu)} \cdot 
		\diff u \diff v $ concisely.
		
		\par As a result, by using the probability generating functional of the Poisson point process on the cylinder set with intensity $ \lambda_l/\pi $, we obtain
		\begin{align}
		\cL_{I_1}(s)&=\bE_{\Phi_l}\left[\left.\prod_{r_j}\exp\left(-\frac{\mu}{\pi \nu^2}\int_{\bR}\int_{B_0(\nu)}\frac{sp{((r_j+u)^2+(t+v)^2)}^{-\frac{\alpha}{2}}}{1+sp{((r_j+u)^2+(t+v)^2)}^{-\frac{\alpha}{2}}}\diff u\diff v\diff t\right)\right.\right]\nnb\\
		&=\exp\left(-\lambda_l\int_{\bR}1-e^{-\frac{\mu}{\pi \nu^2}\int_{\bR}\int_{B_0(\nu)}\frac{sp{((r+u)^2+(t+v)^2)}^{-\alpha/2}}{1+sp{((r+u)^2+(t+v)^2)}^{-\alpha/2}}\diff u\diff v\diff t}\diff r\right)\label{13}.
\end{align}
\par In a similar way, let $ \vec{\kappa_0} $ denotes the unit vector of the typical line.  Then, the locations of the points on the typical line is given by $t_k\vec{\kappa_0} $. As above, consider $ \vec{\kappa_0}^{\perp} $; a unit vector orthogonal to $ \vec{\kappa_0} $. Then, we can write $ \mathbf{u}= u\vec{\kappa_0} + v\vec{\kappa_0}^{\perp}$. The Laplace transform of the interference from the roadside devices  associated with vehicles on the typical line is given by 
\begin{align}
	\cL_{I_2}(s)&=\bE_{\phi(0)}\left[\left.\prod_{X_i\in\phi(0)}\int_{B_0(\nu)}\frac{1}{\pi \nu^2}\frac{1}{1+sp{\|X_i+\mathbf{u}\|}^{-\alpha}}\diff \mathbf{u}\right.\right]\\
	&=\bE_{\phi(0)}\left[\left.\prod_{t_k\in\phi(0)}\int_{B_{0}(\nu)}\frac{1}{\pi \nu^2}\frac{1}{1+sp{\|t_k\vec{\kappa_0}+u\vec{\kappa_0} + v\vec{\kappa_0}^{\perp}\|}^{-{\alpha}}}\diff \mathbf{u}\right.\right]\nnb\\
&=\bE_{\phi(0)}\left[\left.\prod_{t_k\in\phi(0)}\frac{1}{\pi \nu^2}\int_{B_{0}(\nu)}\frac{1}{1+sp{\|(t_k+u)\vec{\kappa_0}+v\vec{\kappa_0}^{\perp}\|}^{-{\alpha}}}\diff {u}\diff v\right.\right]\nnb\\
&=\bE_{\phi(0)}\left[\left.\prod_{t_j\in\phi(0)}\frac{1}{\pi \nu^2}\int_{B_{0}(\nu)}\frac{1}{1+sp{((t_k+u)^2+v^2)}^{-\frac{\alpha}{2}}}\diff {u}\diff v\right.\right]\nnb\\
	&=\exp\left(-\frac{\mu}{\pi \nu^2} \int_{\bR}\int_{B_{0}(\nu)}\frac{sp{((t+u)^2+v^2)}^{-\frac{\alpha}{2}}}{1+sp{((t+u)^2+v^2)}^{-\frac{\alpha}{2}}}\diff u\diff v \diff t\right)\label{12},
\end{align}
where we use the probability generating function on the Poisson point process with intensity $ \mu $. Notice that Eq. \eqref{12} is not a function of the angle of the typical line. 
\par Finally, from the independence of random variables $ I_1 $ and $ I_2 $, we multiply Eqs. \eqref{13} and  \eqref{12} to obtain the complete formula for the Laplace transform of interference. 
\end{IEEEproof}
\begin{remark}
In the proposed vehicular architecture, the interference power of the typical vehicle is given by Eq. \eqref{6}. In contrast, the interference seen by a typical point in the plane does not follow the same distribution. Specifically, from the perspective of an arbitrary located point, it is almost surely not on any road. Therefore, its Laplace transform would be given by Eq. \eqref{13}. Consequently, the typical vehicle experiences an additional interference from the devices on the typical road compared to a randomly located point in space. A similar phenomenon was also discussed in \cite{choi2017analytical}. 
\end{remark}

\begin{figure}
	\centering
	\includegraphics[width=.7\linewidth]{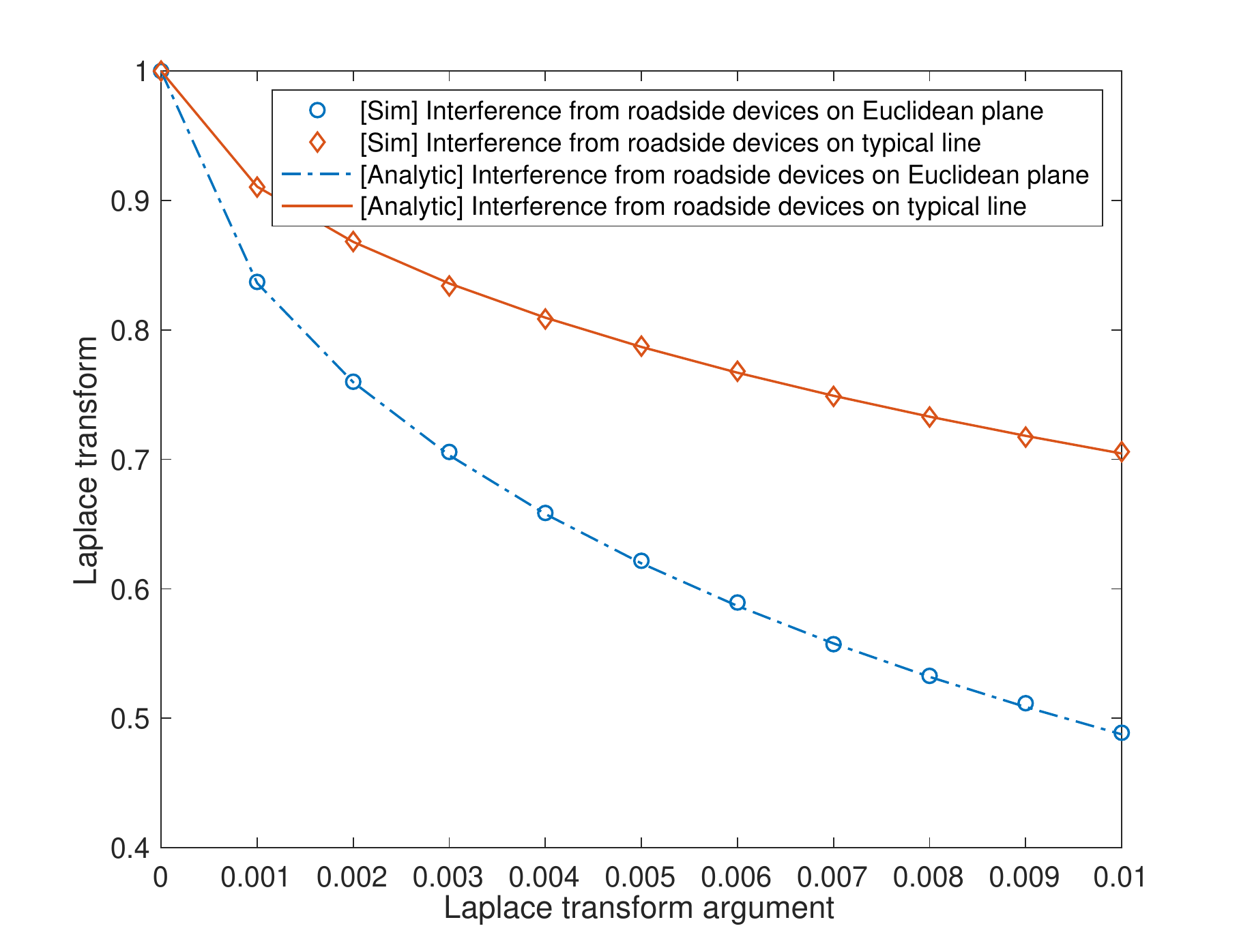}
	\caption{Illustration of the Laplace transforms of the interference: analytic vs. simulation.}
	\label{fig:numericalint}
\end{figure}

\par Fig. \ref{fig:numericalint} illustrates the Laplace transforms of the interference, evaluated by Monte-Carlo simulations and derived by Theorem \ref{T:1}, respectively. By comparing marks and lines, we find that the interference integral expression in Theorem \ref{T:1} exactly matches the simulation result. For the computation, we use $ p=0.01, \alpha=3, \lambda_l=\mu=5,$ and $ \nu=0.1 $. The $ x$-axis is the Laplace transform argument.


\subsection{Coverage Probability}

\begin{theorem}\label{T:1}
	The SIR coverage probability of the typical vehicle is given by 
	\begin{align}
	p_c(\tau)=\int_{0}^{\nu}&\exp\left(-\lambda_l\int_{\bR}1-e^{-\frac{\mu}{\pi \nu^2}\int_{\bR}\int_{B_0(\nu)}\frac{\tau u^{\alpha}{((r+u)^2+(t+v)^2)}^{-\alpha/2}}{1+\tau u^{\alpha}{((r+u)^2+(t+v)^2)}^{-\alpha/2}}\diff u\diff v\diff t}\diff r\right)\nnb\\
	&\exp\left(-{\mu}\int_{\bR}\int_{B_{0}(\nu)}\frac{\tau u^{\alpha}{((t+u)^2+v^2)}^{-\alpha/2}}{1+\tau u^{\alpha}{((t+u)^2+v^2)}^{-\alpha/2}}\diff u\diff v \diff t\right)\frac{2u}{ \nu^2}\diff u.\label{eq:T1}
	\end{align}
\end{theorem}
\begin{IEEEproof}
Denote by $ U $, the location of the roadside device associated with the typical vehicle at the origin. Then, the SIR coverage probability of the typical vehicle is given by 
\begin{align}
p_c(\tau)&=\bP_{\Phi}^{0}\left(\frac{pH{\|U\|}^{-\alpha}}{\sum\limits_{X_i\in\Phi+\delta_0} p H {\|X_i+U_i\|}^{-\alpha}}>\tau\right)\nnb\\
&=\bP_{\Phi}^0\left(\frac{pH{\|U\|}^{-\alpha}}{\sum\limits_{X_i\in\Phi^{!0}}p H {\|X_i+U_i\|}^{-\alpha}}>\tau\right)\nnb\\
	&\ea\bP\left(H>\tau{\|U\|}^{\alpha}p^{-1}I\right)\eb\int_{0}^{\nu}\frac{2u}{ \nu^2}\bE\left[e^{-\tau u^\alpha p^{-1} I}\right]\diff u,\nnb
\end{align}
where $ n $ denotes the noise power. We obtain (a) from Slivnyak's theorem and (b) by using the density of $ \|U\| $  as $ 2u/\nu^2 $ as for $ 0\leq u \leq \nu. $ 
The proof is completed by using Eq. \eqref{6}.
\end{IEEEproof}

\begin{figure}
	\centering
	\includegraphics[width=.7\linewidth]{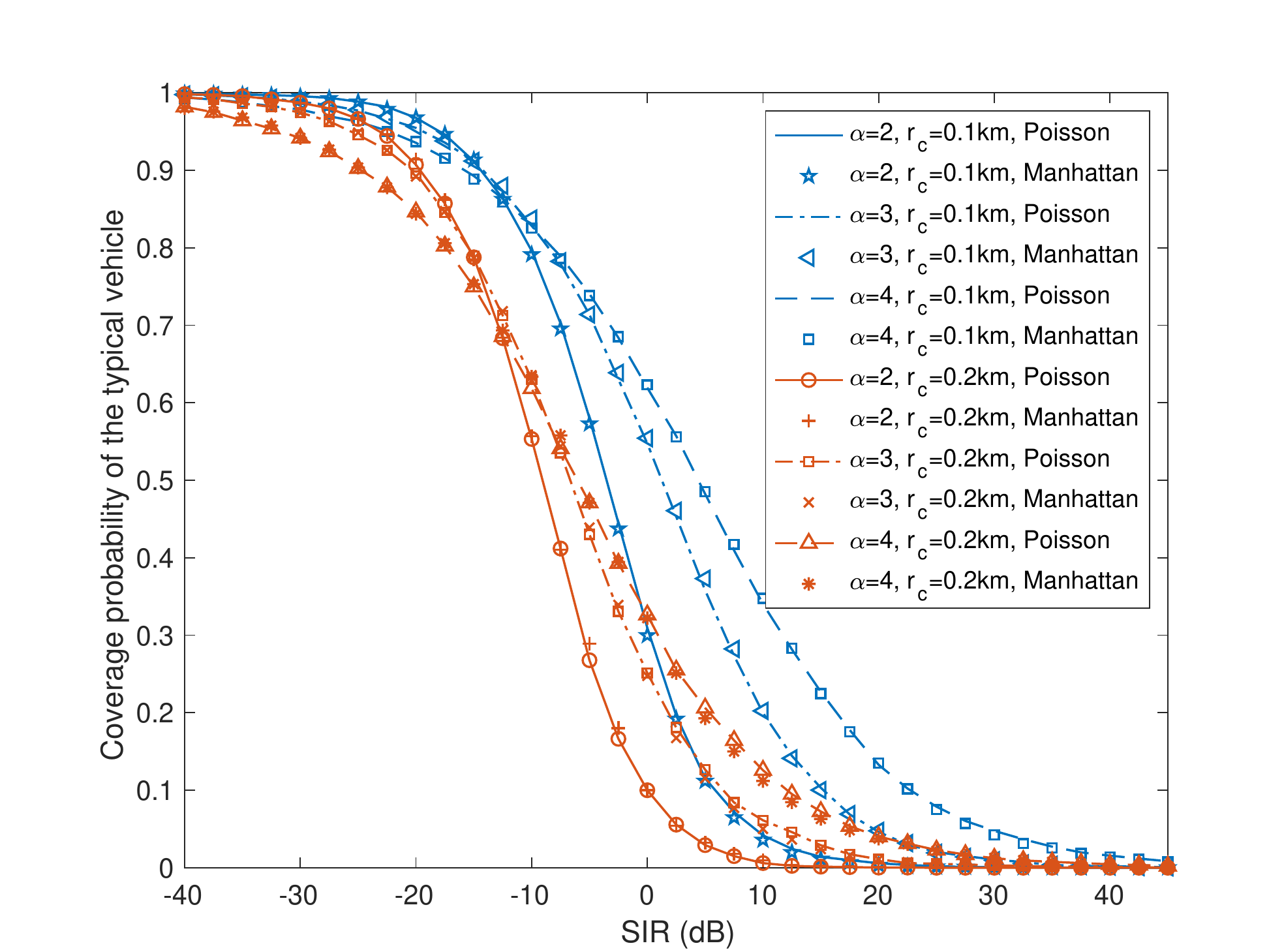}
	\caption{The coverage probability of the typical vehicle when  $ \lambda_l=3/\text{km}, $ $ \mu=3/\text{km}, $ $ \nu=\{0.1,0.2\}/\text{km}, $ and $ \alpha=\{2,3,4\}. $}
	\label{fig:coverage}
\end{figure}

 Fig \ref{fig:coverage} illustrates the coverage probability of the typical vehicle. For the considered parameters, the figure exhibits two trends with respect to parameter $ \alpha $ and $ \nu $; (1) a higher path loss exponent provides better coverage due to better spatial separation of the interference for the same topology of vehicles and devices; (2) a smaller disk yields a better coverage performance mainly because it implies that the roadside devices are closer to their corresponding vehicles. These observations substantiate the claim that the proposed architecture ensures a better SIR in a dense urban scenario where the path loss is larger and the distances from the roadside devices to roads are shorter.

\subsection{Area Spectral Efficiency}
The area spectral efficiency is defined by the product of the achievable rate of the typical link and the density of vehicles. Consequently, the area spectral efficiency of the network is interpreted as the spatial average of the achievable throughput in a unit area. 

\begin{figure}
	\centering
	\includegraphics[width=0.7\linewidth]{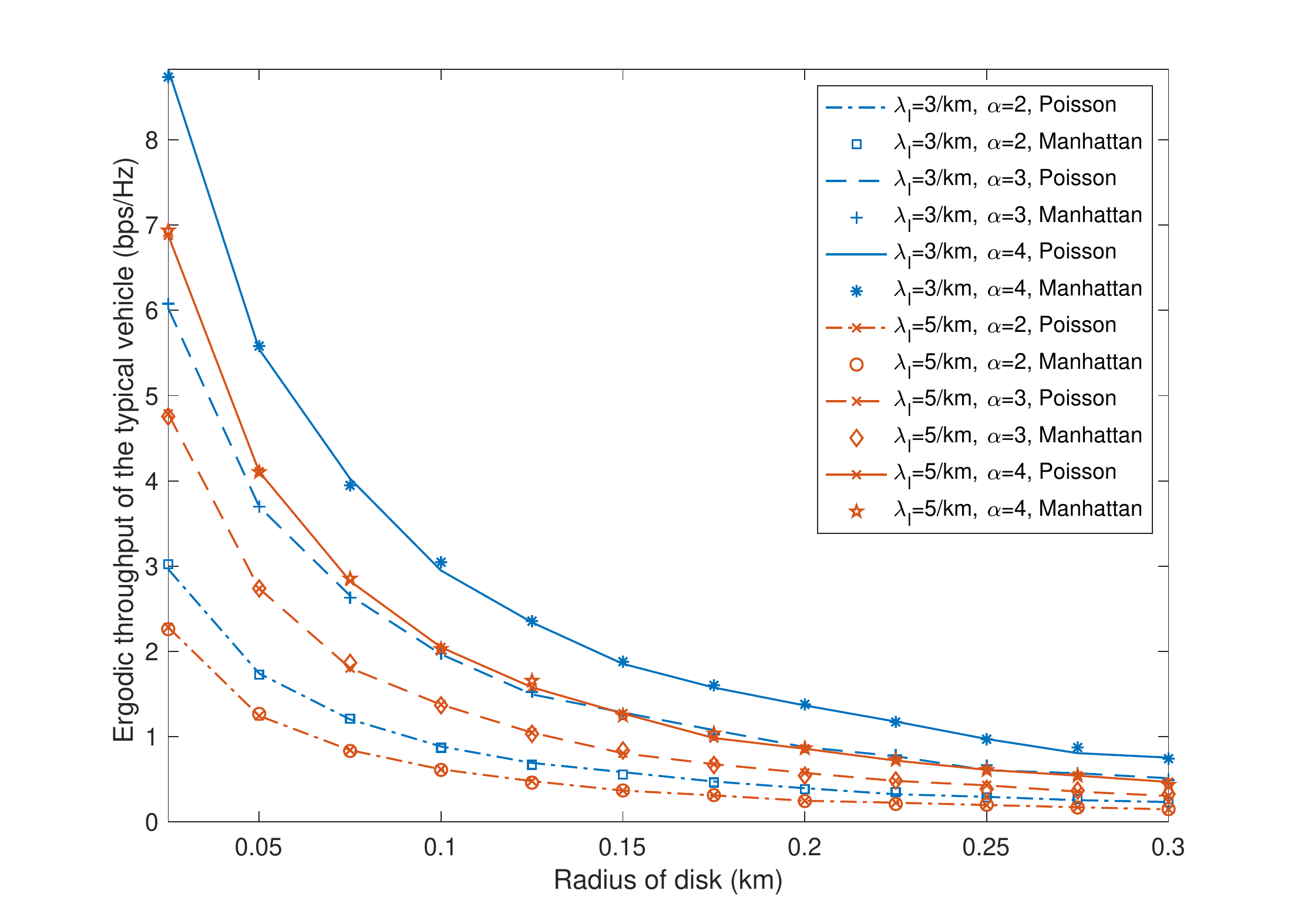}
	\caption{Illustration of the ergodic throughput of the typical vehicle at the origin when $ \mu=3/\text{km} $.}
	\label{fig:asepl0}
\end{figure}

\begin{theorem}\label{T:2}The area spectral efficiency of the proposed architecture is given by 
	\begin{equation}
		\text{ASE}= \lambda_l\mu\int_{0}^{\infty}\int_{0}^{\nu}\frac{2pr^{1-\alpha}}{\nu^2(1+z p r^{-\alpha})}\cL_{I}(z)\diff r\diff z,
	\end{equation}
	where $ \cL_{I}(z) $ is given in Theorem \ref{T:1}.
\end{theorem}
\begin{IEEEproof}
 To obtain the ergodic throughput of the typical link we use the following expression in \cite{hamdi2010useful}. For two independent random variables $ X>0$ and $Y>0 $,  we have 
\begin{equation*}
\bE\left[\log_2\left(1+\frac{X}{Y}\right)\right]=\int_{0}^{\infty}z^{-1}\left(1-\bE\left[e^{-Xz}\right]\right)\bE\left[e^{-Yz}\right]\diff z.
\end{equation*}
The area spectral efficiency is given by the product of the density and the ergodic throughput. We have   
\begin{align}
&\lambda_l\mu\bE_{\Phi}^{0}\left[\log_2\left(1+\frac{pH\|U\|^{-\alpha}}{\sum_{X_i\in\Phi^{!0}}pH{\|X_i+U_i\|}^{-\alpha}}\right)\right]\nnb\\
&\ea\lambda_l\mu\int_{0}^{\infty}z^{-1}\left(1-\bE_{H,U}\left[e^{-zpH\|U\|^{-\alpha}}\right]\right)\bE_{\Phi}^{0}\left[e^{-zI}\right]\diff z\nnb\\
&\eb\lambda_l\mu\int_{0}^{\infty}z^{-1}\left(1-\int_{0}^{\nu}\frac{2r}{\nu^2(1+z p r^{-\alpha})}\diff r\right)\bE_{\Phi}^{0}\left[e^{-zI}\right]\diff z\nnb\\
&=\lambda_l\mu\int_{0}^{\infty}{z}^{-1}\left(\int_{0}^{\nu}\frac{2zpr^{1-\alpha}}{\nu^2(1+z p r^{-\alpha})}\diff r\right)\bE_{\Phi}^{0}\left[e^{-zI}\right]\diff z\nnb\\
&=\lambda_l\mu\int_{0}^{\infty}\int_{0}^{\nu}\frac{2pr^{1-\alpha}}{\nu^2(1+z p r^{-\alpha})}\cL_{I}(z)\diff r\diff z\label{18}.
\end{align}
To derive (a), we use the interference seen by the typical vehicle obtained by Theorem \ref{T:1}. To get (b), we use that $ U$ is randomly distributed in disk $ B_0(\nu) $ and the density of its norm $ \|U\| $. Finally, applying Eq. \eqref{6} to \eqref{18} completes the proof. 
\end{IEEEproof}
Fig. \ref{fig:asepl0} illustrates the ergodic throughput, i.e., the achievable rate of the typical vehicle. For the considered parameters, a higher path loss exponent and/or a lower road intensity provide a better link throughput. Moreover, as the size of coverage disk increases, the link throughput monotonically decreases because the mean distances from vehicles to their associated roadside devices also increase.


\section{Performance Analysis in the Time Domain}\label{S:4}
We analyze the evolution of the coverage disks with respect to time by deriving the area fraction and latency. 

\subsection{Area Fraction of Coverage Disks}
Recall that for a time $ t $, the union of the coverage disks is given by 
\begin{align*}
S(t)=&\bigcup_{X\in\Phi(t)}B_{X}(\nu).
\end{align*}
\begin{theorem}\label{T:3}
	For any time $ t, $ $ S(t) $ is time and motion invariant. Furthermore, the area fraction of $ S(t) $ is given by 
	\begin{equation}
	\text{AF}(S(t))=1-\exp\left(-2\lambda_l\int_{0}^{\nu}1-e^{-2\mu\sqrt{\nu^2-{u}^2}}\diff u\right).\label{22}
	\end{equation}
		Moreover, the area fraction of the cumulative coverage disks up to time $ t $ is given by  
	\begin{align}
	\text{AF}({\bar{S}}(t))=1-\exp\left(-2\lambda_l\int_{0}^{\nu}1-\exp\left(-2\mu (vt + \sqrt{\nu^2-r^2})\right)\diff r\right)\label{28}.
	\end{align} 
	In addition, its limit is given by 
	\begin{equation}\label{eq:25}
	\text{AF}({\bar{S}}(\infty))=1-\exp\left(-2\lambda_l\nu\right).
	\end{equation}
\end{theorem}
\begin{IEEEproof}
We first show that $ S(t) $ is time invariant. By a slight abuse of notation, let $ \phi(r_i,\theta_i,t=0) $ denote at the Poisson point process on line $ (r_i,\theta_i,t=0) $ at time zero. Then, we have 
\begin{equation}
S(t=0)=\bigcup_{X\in\Phi(t=0)}B_X(\nu) =\bigcup_{(r_i,\theta_i)\in\Psi(t=0)}\left({\bigcup_{X\in\phi{(r_i,\theta_i,t=0)}} B_{X}(\nu)}\right),
\end{equation} where one can interpret the set $ {\bigcup_{X\in\phi{(r_i,\theta_i,t=0)}} B_{X_{i,j}}(\nu)} $ as the Boolean model \cite[Chap.3]{baccelli2010stochastic} of finite radius balls centered on the Poisson point process $ \phi(r_i,\theta_i,t=0) $ at time zero. Notice that every vehicle is assumed to choose its moving directions, according to an independent and idential coin toss at time zero. Therefore,  the point process at time $t=\delta $ is still a Poisson point process on the line with the same intensity. Furthermore, since the distributions of the centroids of the Boolean models are the same for time $ 0$ and $t $, we can write 
\begin{equation}
	\bigcup_{X\in\phi{(r_i,\theta_i,t=0)}} B_X(\nu)\stackrel{d}{=}\bigcup_{X\in\phi{(r_i,\theta_i,t=\Delta)}} B_X(\nu),
\end{equation}
where $ \stackrel{d}{=} $ denotes equality in distribution. 
Because the Poisson lines are time-invariant, we have 
\begin{align}
	S(t=\Delta )\stackrel{d}{=}S\label{23},
\end{align}
where $ S $ denotes the union of the coverage disks at time zero.
\par In order to show the planar motion invariance of $ S, $ we use \cite[Prop. 4.3]{molchanov2005theory}; the random closed set $ S $ is motion invariant if and only if its capacity functional, $ \cT_S(K):=\bP(S\cap K \neq \emptyset)$ for all compact set $ K, $ is motion invariant \cite{chiu2013stochastic}. The capacity functional of $ S $ is given by
\begin{align}
	\mathcal{T}_S(K)&=1-\bP(\text{no point of } \Phi \text{ in } B_0(\nu)).\nnb
	\end{align}
	Furthermore, we also have  
	\begin{align}
	\mathcal{T}_S(K+x)&=1-\bP(\text{no point of } S_{x}\Phi \text{ in } B_x(\nu))= 1-\bP(\text{no point of }  \Phi  \text{ in } B_0(\nu)),\nnb
\end{align}
where $ S_x\Phi $ is the translation of $ \Phi $ by $ x\in\bR^2. $
We obtain the last expression because the vehicle point process $ \Phi $ (the grains of the Boolean model) is a motion invariant point process\cite{choi2018poisson}. Since $ \cT_S(K)=\cT_S(K+x) $, the random closed set $ S $  is motion invariant. 

\par Leveraging the invariance property of $ S(t) $, the area fraction of $ S(t) $ is 
\begin{align*}
\text{AF}(S(t))&=\lim_{r\to\infty}\frac{\bE\left[\ell_2(S(t)\cap B(r))\right]}{\ell_2(B(r))}\\
&\ea\lim_{r\to\infty}\frac{\bE\left[\ell_2(S\cap B(r))\right]}{\ell_2(B(r))}\\
&\eb\lim_{r\to\infty}\frac{\bE\left[\bigintssss_{B(r)}\ind_{x\in S}\diff x\right]}{\bigintssss_{B(r)} \ind \diff x}\\
&\ec\lim_{r\to\infty}\frac{\bigintssss_{B(r)}\bE\left[\ind_{x\in S}\right]\diff x}{\bigintssss_{B(r)} \ind \diff x}\ed\bE[\ind_{0\in S}],
\end{align*}
where (a) is obtained by time invariance of $ S $ and (b) follows from the fact that the area of a set is given by the Lebesgue integral of the indicator function of the set. We obtain (c) from Fubini's theorem and (d) from the stationarity of $ S $, respectively. Therefore, the area fraction is  
\begin{align}
\bE[\ind_{0\in S}]=\bP(0\in S)\ee\bP(\min_{X_i\in\Phi}\|X_i\|\leq \nu)=1-\bE\left[\prod_{X_i\in\Phi}\ind_{\|X_i\|> \nu}\right],\label{24}
\end{align}
where we obtain (e) because the probability that $ S $ containing the origin is equivalent to the probability that $ B_{\min_{X_i\in\Phi}\|X_i\|}(\nu), $ the disk centered at the $ \min_{X\in\Phi}\|X\| $ includes the origin. Moreover
\begin{align}
\bE\left[\prod_{X_i\in\Phi} \ind_{\|X_i\|>\nu}\right]	&\ef\bE_{\Psi}\left[\prod_{r_i}\bE_{\phi(r_i,\theta_i)}\left[\left.\prod_{X\in\phi(r_i,\theta_i)}\mathbbm{1}_{\|X\|>\nu}\right|\Psi\right]\right]\nnb\\
&\eg\bE_{\Psi}\left[\prod_{r_i}\bP\left(\|X_{i,0}\|\wedge \|X_{i,1}\|> \sqrt{\nu^2-r_i^2}\right)\right]\nnb\\
&\eh\bE_{\Psi}\left[\prod_{r_i}\exp\left(-2 \mu \sqrt{\nu^2-|r_i|^2}\right)\mathbbm{1}_{-r<r_i<r}\right]\nnb\\
&=\exp\left(-2\lambda_l\int_{0}^{\nu}1-e^{-2\mu\sqrt{\nu^2-{u}^2}}\diff u\right),\label{25}
\end{align}
where $ x\wedge y $ denotes the minimum of $ x $ and $ y. $
We have (f) by conditioning on the Poisson line process $ \Psi $ and by representing each Cox point $ X_k $ as $ X_{i,j}, $ where index $ i $ denotes the line $ l(r_i,\theta_i) $ on which $ X_{k} $ is located and index $ j $ counts the point on this line. Equality (g) is obtained by denoting the two nearest point---with respect to the closest point of the line $ i $ to the origin---on each side by $ X_{i,0} $ and $ X_{i,1} $, respectively. We obtain (h) from the distribution function of the exponential random variable. Applying the Laplace transform of the Poisson point process \cite{chiu2013stochastic} gives Eq. \eqref{25}. Finally combining Eq. \eqref{25} into Eq. \eqref{24} completes the proof. 

\par Now, let us focus on the cumulative coverage disks $ \bar{S}(t) $. The area fraction is given by 
\begin{align*}
\text{AF}({\bar{S}}(t))&=\lim_{r\to\infty}\frac{\bE\left[\ell_2(\bar{S}(t)\cap B(r))\right]}{\ell_2(B(r))}\\
&=\lim_{r\to\infty}\frac{\bE\left[\bigintssss_{B_0(r)}\ind_{x\in \bar{S}(t)}\diff x\right]}{\bigintssss_{B_0(r)} \ind \diff x}\\
&=\lim_{r\to\infty}\frac{\bigintssss_{B_0(r)}\bE\left[\ind_{x\in \bar{S}(t)}\right]\diff x}{\bigintssss_{B_0(r)} \ind \diff x}\ed\bE[\ind_{0\in \bar{S}(t)}].
\end{align*}
Therefore, we have $ \text{AF}({\bar{S}}(t))=\bP(0\in\bar{S}(t))=1-\bP(0\notin\bar{S}(t)), $ where  $ 0\notin \bar{S}(t) $ means that the origin is not an element of set $ \bar{S}(t) $. Then, in order to have the set $ \bar{S}(t) $ not containing the origin at time $ t $, the following two conditions should be satisfied: (1) the distances from the origin to the lines are greater than $ \nu, $ or (2) for the lines $ (r_i,\theta_i) $ whose distances are less than $ \nu, $ all of their points satisfy $ \|X_{i,j}-0_i\|>  \sqrt{\nu^2-r_i^2}+ v t, $ where $ X_{i,j} $ is the $ j$-th point on the Poisson line $ r_i $ and $ 0_i$ is the point on line $ i $ closest to the origin. As a result, we have 
\begin{align}
\text{AF}({\bar{S}}(t))&=1-\bP(0\notin \bar{S}(t))\nnnl
&= 1-\bE\left[\prod_{r_i}\bE\left[\prod_{X\in\phi(r_i,\theta_i)}\ind_{X>vt+\sqrt{\nu^2-r_i^2}}\right]\right]\nnnl
&{=} 1-\bE\left[\prod_{r_i}\exp\left(-2\mu (vt+\sqrt{\nu^2-r_i^2})\right)\right]\nnb\\
&\stackrel{\text{(i)}}{=} 1-\exp\left(-2\lambda_l\int_{0}^{\nu}1-\exp\left(-2\mu\left(vt+\sqrt{\nu^2-r^2}\right)\right)\diff r\right).\nnb
\end{align}
The limit value of the area fraction is obtained by taking $ t=\infty $.
\end{IEEEproof}

Theorem \ref{T:4} shows that the area fractions of $ S(0) $ and $ S(t) $ have the same distribution; in particular, the areas covered at time $ 0 $ and at time $ t $ are the same on average. In Figs. \ref{fig:proposednetwork} and \ref{fig:proposednetwork_manhattan}, the area fraction can be interpreted as the mean area of the shaded region, divided by the total area. Eq. \eqref{22} gives the area fraction as a function of the parameters. It shows that the area fraction is increasing with the radius of the coverage disk $ \nu, $ the intensity of vehicles $ \mu$, and the density of lines $ \lambda_l $.

\begin{figure}
	\centering
	\includegraphics[width=0.7\linewidth]{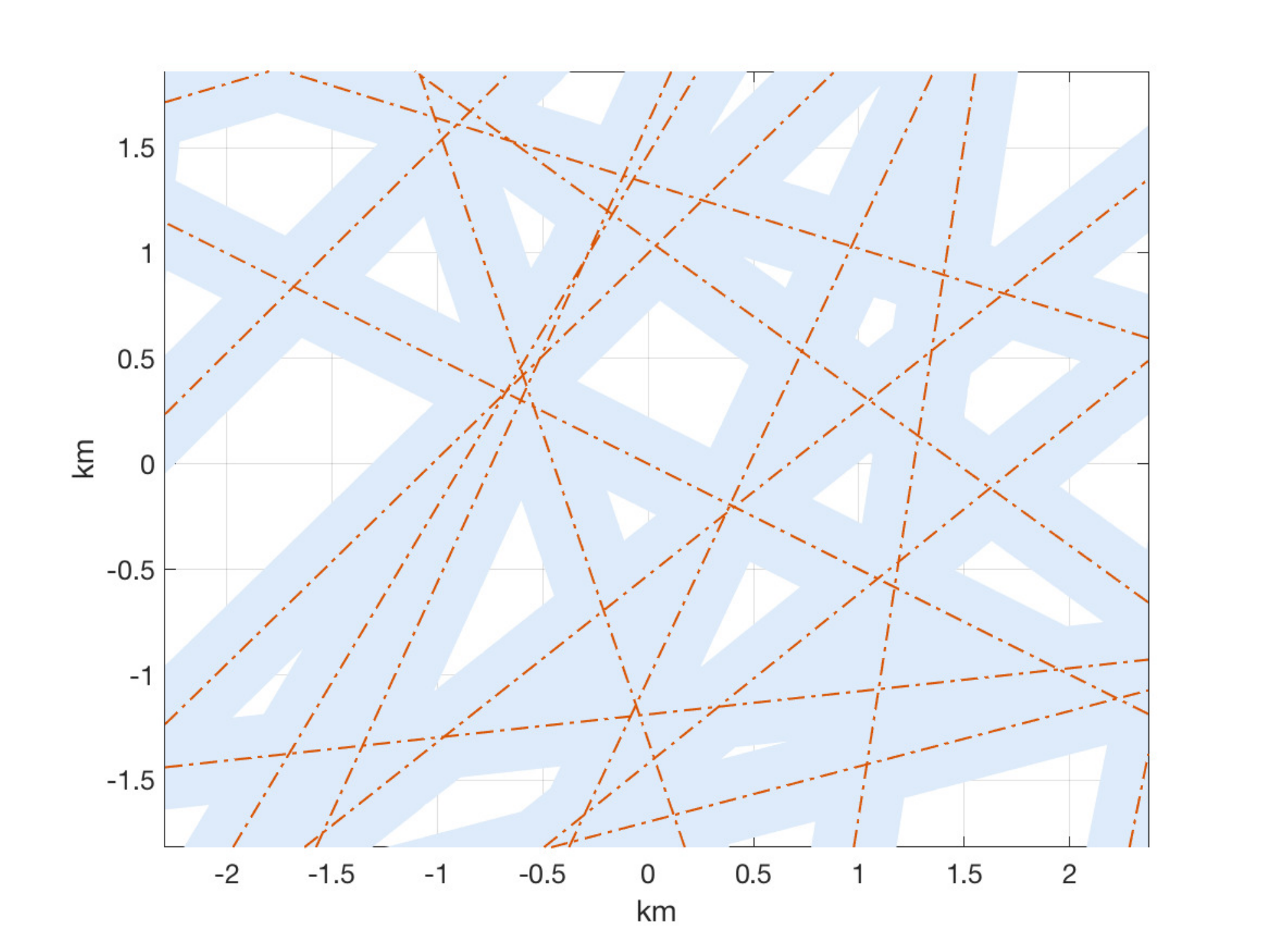}
	\caption{For parameters $ v=100\text{km/h}, $$ \lambda_l=4/\text{km} $ and $ \mu=5/\text{km}, $ the cumulative coverage disks at $ t=1000 $ seconds is illustrated as shaded area. The dashed lines are given to indicate Poisson roads. }
	\label{fig:areafractioninf}
\end{figure}

\begin{figure}
	\centering
	\includegraphics[width=0.7\linewidth]{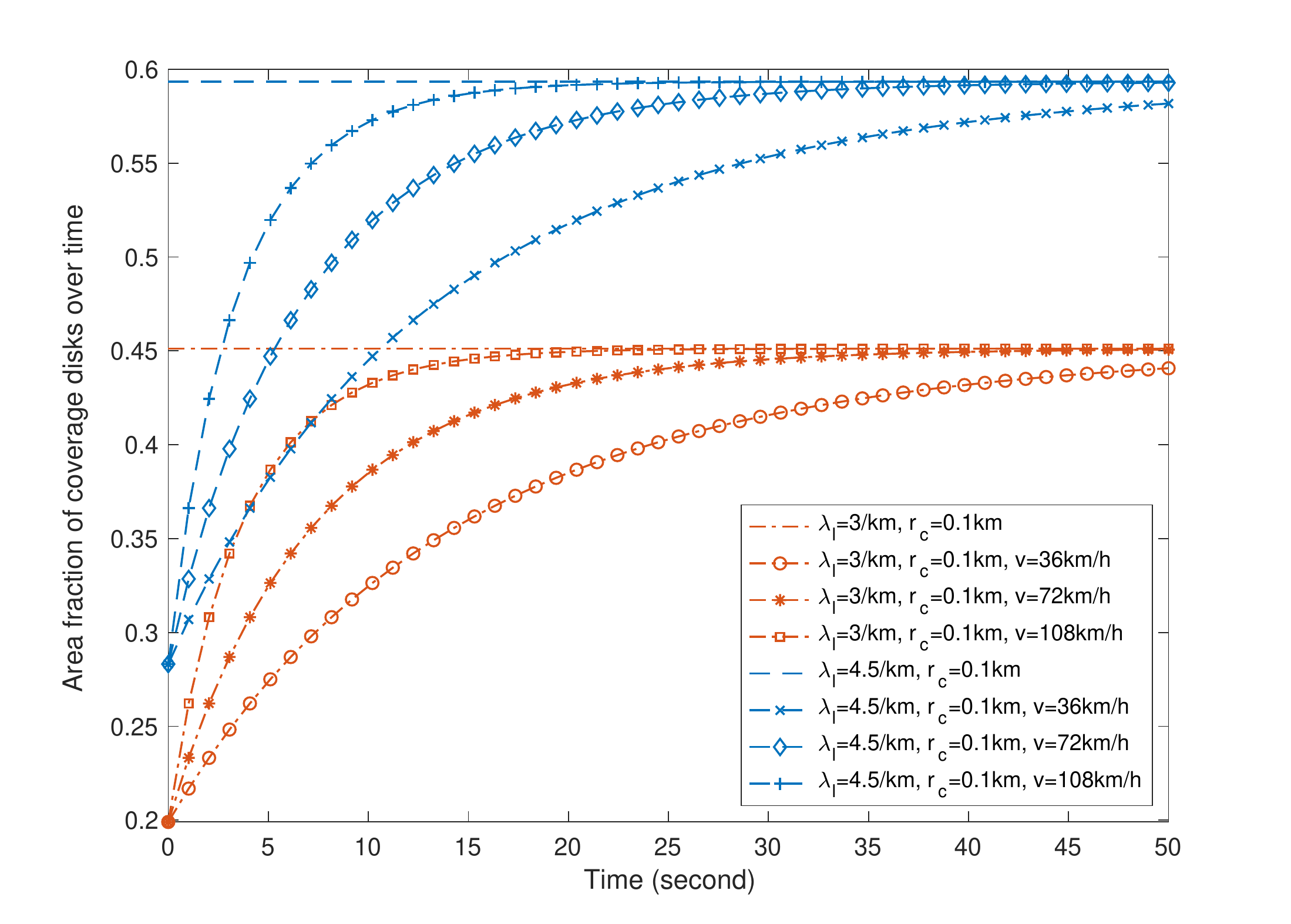}
	\caption{Illustration of the area fractions of the coverage disks over time and their limits. }
	\label{fig:nsrr2}
\end{figure}
Now, let us discuss $ \text{AF}(\bar{S}(t)) $: (1) it is increasing with respect to time $ t; $ (2)  it is a function of the radius of coverage disk $ \nu $ and the intensity $ \lambda_l $ of the line process only. Fig \ref{fig:nsrr2} depicts the behaviors of $ \text{AF}({\bar{S}}(t)) $ with respect to time, for two sets of parameters. Note that as time tends to infinity, $ \text{AF}({\bar{S}}(t)) $ tends to $ \text{AF}({\bar{S}}(\infty)) $ specified by Eq. \eqref{eq:25}. For moderate speeds $ v=\{36,72,108\}\text{km}/\text{h}, $ the limiting values $ \text{AF}({\bar{S}}(\infty)) $ are achieved in less than $ 60 $ seconds. Notice that the limit is increasing with respect to the density of roads and the radius of the coverage disk. Eqs. \eqref{28} and \eqref{eq:25} show that long-term network properties are affected by parameter changes, as we have seen in the short-term network performance.

\begin{figure}
	\centering
	\includegraphics[width=0.7\linewidth]{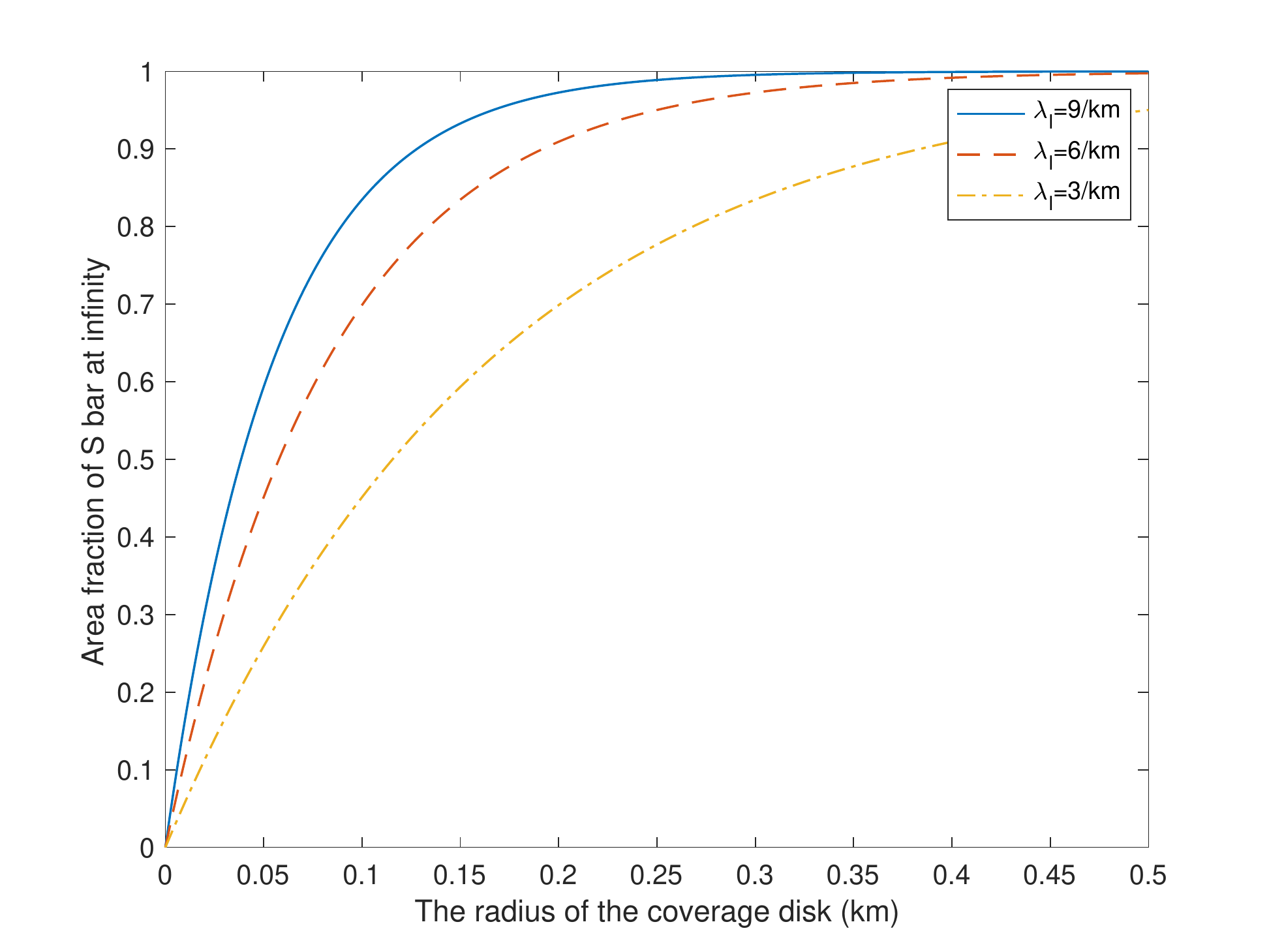}
	\caption{The area fractions of the limiting values $ {\bar{S}(\infty)} $ for dense urban, suburban, and rural scenarios. }
	\label{fig:afsinf}
\end{figure}

\begin{remark}
	Fig. \ref{fig:afsinf} illustrates the area fraction of the covered area in Eq. \eqref{eq:25} with respect to the radius of the coverage disk, for three different cases: (1) urban, (2) suburban, and (3) rural, which are devised according to the density of roads and vehicles. For the urban case, the road density is high $ \lambda_l=9/\text{km} $; for the suburban area, it is moderate, $ \lambda_l=6/\text{km} $; and for the rural area case, it is low, $ \lambda_l=3/\text{km} $.  For all cases, the limits tend to one as the radius of the coverage disk tends to infinity. Furthermore, the urban area case dominates both the suburban and rural cases. This phenomenon indicates that a smaller disk may suffice to cover a significant part of the Euclidean plane in the urban case. For instance, when the radius of the disk is $ 100 $ meters---the typical transmission range of devices  with limited power source\cite{gubbi2013internet}---about 80\% of the entire plane is covered in the urban case whereas only 30\% is covered in the rural case. 
\end{remark}

The following examples shed light on Theorem \ref{T:3} in different mobility and vehicle deployment. 
\begin{example}
We have assumed that vehicles move at a constant speed $ v $ and that their trajectories strictly follow roads. An extension of the proposed model could feature a randomized-speed model where each vehicle initially determines its speed according to an independent and identical distribution such as the $ \text{Normal}(v,\sigma^2) $ and maintains its speed. By the displacement property of the Poisson point process, the locations of the vehicles after time $ t $ on each line is again given by a Poisson point process with intensity $ \mu. $ Therefore, the area fraction is still given by Eqs. \eqref{22} and \eqref{28}.
\end{example}


\subsection{Latency}
In this paper, the network latency is characterized as the average time for the typical device to be located inside a coverage disk. \footnote{When the typical device is not contained by $ \bar{S}(\infty), $ the waiting time this case is defined as infinity.} For simplicity, we define the network latency conditioning on the fact that the typical device is contained by set $ \bar{S}(\infty). $ See Fig. \ref{fig:areafractioninf} for the illustration of $ \bar{S}(\infty). $ For Poisson distributed devices, the probability of being contained in $ \bar{S}(\infty) $ is equal to  $ 1-\exp(-2\lambda_l\nu). $ 
\begin{figure}
	\centering
	\includegraphics[width=0.7\linewidth]{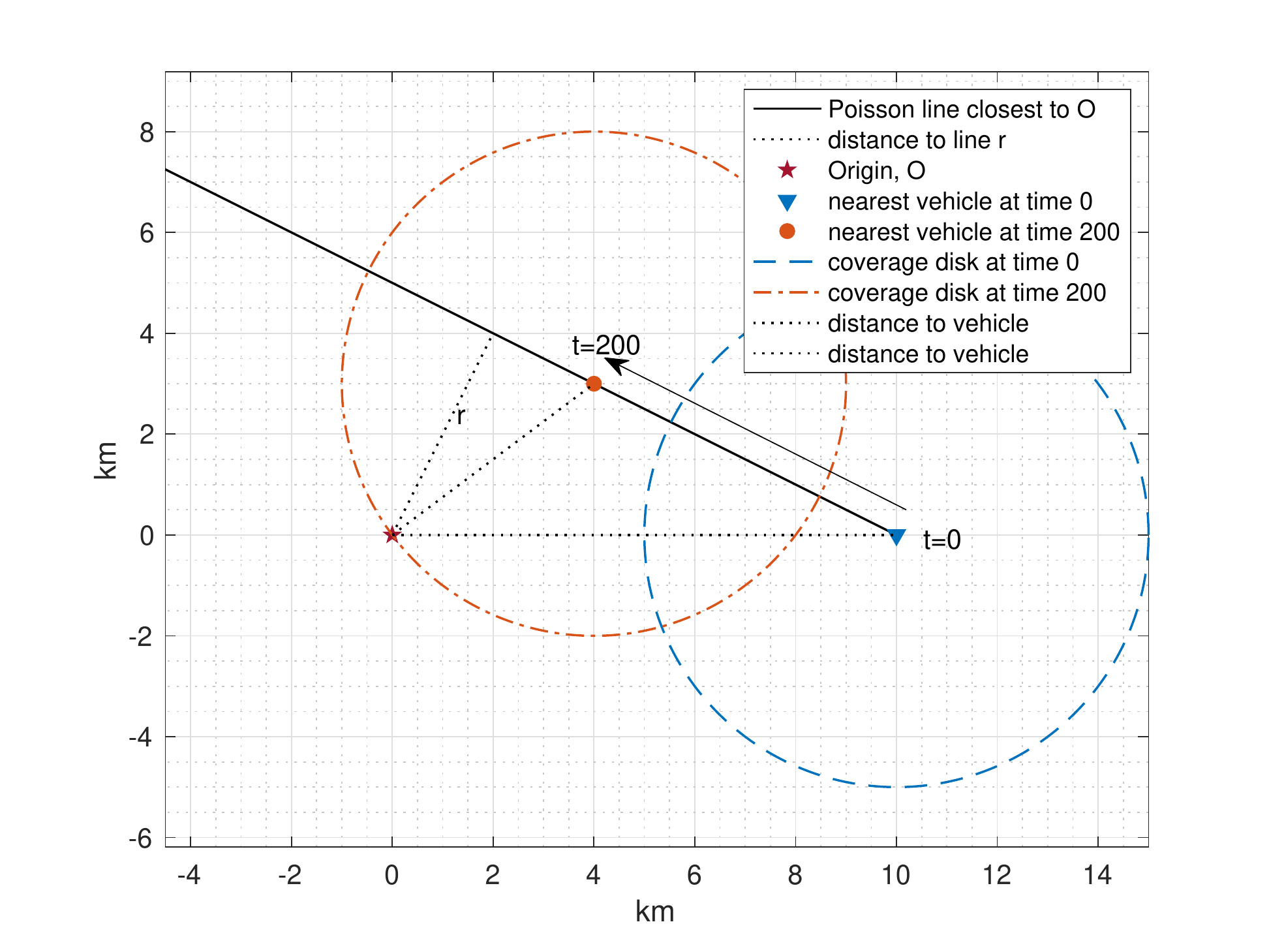}
	\caption{Changes of coverage area from $ t=0\text{ sec} $  to $t= 200 \text{ sec}. $ The speed of vehicle is around $ 120 $ km/h. }
	\label{fig:illustrationline}
\end{figure}

\begin{theorem}\label{T:4}
The network latency is given by 
	\begin{equation}\label{eq:T:4}
		\int_{0}^{\infty}\exp\left(-2\lambda_l\int_{0}^{\nu}1-e^{-2\mu \left(\sqrt{\nu^2-u^2}\right)}\diff u\right)\diff w.
	\end{equation}
	
	
\end{theorem} 
\begin{IEEEproof} We can write the distribution of latency as 
	\begin{align}
		\bP(W>w|0\in\bar{S}(\infty))&=\bE\left[\prod_{r_i}\bE_{\phi(r_i,\theta_i)}\left[{\prod_{T_j}\ind_{|T_{j}|>vw+\sqrt{\nu^2-r_i^2}}}\right]\right]\nnb\\
		&=\bE\left[\prod_{r_j}\exp\left(-2\mu \left(\sqrt{\nu^2-r_i^2}+vw\right)\right)\right]\nnb\\
		&=\exp\left(-2\lambda_l\int_{0}^{\nu}1-e^{-2\mu \left(\sqrt{\nu^2-u^2}+vw\right)}\diff u\right)\label{244},
	\end{align}
	for all $ w\geq 0 $, where we have used the same technique as in \eqref{25}. In fact, $ W=0 $ if the typical device is contained at time zero. The probability is equal to the area fraction of set $ S $ and is given by 
	\begin{equation}
		\bP(W=0|0\in\bar{S}(\infty))= 1-\exp\left(-2\lambda_l\int_{0}^{\nu}1-e^{-2\mu \sqrt{\nu^2-u^2}}\diff u\right).
	\end{equation}
	\par Since the expectation of random variable $ X\geq 0 $ is $ \int_{0}^{\infty}\bP(X>x)\diff x $, the network latency is 
	\begin{equation}
\bE[W|0\in\bar{S}(\infty)]=\int_{0}^{\infty}\exp\left(-2\lambda_l\int_{0}^{\nu}1-e^{-2\mu \left(\sqrt{\nu^2-u^2}+vw\right)}\diff u\right)\diff w.	
		\end{equation}
Note that $ \bP(W=0|0\in\bar{S}(\infty)) $ is nonzero and it corresponds to the area fraction of $ S $. 
\end{IEEEproof}
Fig. \ref{fig:illustrationline} illustrates the network latency using the coverage disk at time instances $ t=0$ and $ 200. $ The typical device is located at the origin and the vehicle on line starts to cover the origin at $ t=200\text{ seconds} $ In this case, the network latency is given as $ 200 \text{ seconds} $. We investigate the mean of latency.
\par 
When devices have delay constraints, Eq. \eqref{eq:T:4} helps one to choose the right size of coverage disk or the right density of vehicles. For instance, if the vehicle density $ \mu $ is too small, the latency might be unacceptably large. 

\section{Discussions}\label{S:5}
We discuss a few trade-off relationships in the proposed network. In particular, some trade-off relations are well captured through a single linear combination of short-term and long-term metrics with respect to the network parameters. We then present a qualitative comparison with a static wireless architecture based on the four metrics we studied. 
\subsection{Trade-offs: Short-term vs. Long-term}
Notice that a trade-off relationship exists between the short-term vs. the long-term performance results; specifically, a trade-off exists between the SIR coverage and the latency with respect to the radius of coverage disk. For instance, if the radius increases, a shorter waiting time is achieved according to Theorem \ref{T:4}; i.e., the latency reduces but the SIR coverage probability of the typical vehicle diminishes according to Theorem \ref{T:1}. This contrasting behaviors occur because the size of the disk is closely related to both the distance to the desired signal device and the total area traveled by the coverage disks. 

\par Another trade-off relationship exists between  coverage (or equivalently rate) and  latency with respect to the density of vehicles $ \mu $. For instance, if the density $ \mu $ is high, there are more vehicles on each road, creating a higher interference seen by the typical vehicle. Consequently, the SIR coverage probability decreases. However, due to the increased number of vehicles, the typical point is more likely to be geometrically covered within a short period of time. 

\par This is not very surprising in the context of delay-tolerant networks where the coverage or capacity are known to be improved at the expense of excessive delays. For instance,  \cite{grossglauser2002mobility} showed that by allowing an infinite delay, one can significantly increase the network throughput. Similar trade-off relationships were investigated under different network topologies in \cite{fall2008dtn,pereira2012delay}. The core mechanism of classical delay-tolerant networks is very simple: {nodes transmit only when they are close; this increases the signal power and improves the network performance.}  In this context, the proposed network is very similar to traditional delay-tolerant network type. However, there exists a noticeable difference in the proposed model. In this paper, the trade-off phenomena can be controlled or even exploited by changing parameters such as  the density $ \mu $ or the radius $ \nu $. For instance, by decreasing the density of vehicles, the coverage or rate of the typical vehicle can be significantly improved.  Similarly, by increasing the radius of the coverage disk, the network latency can be substantially improved. Therefore, one can leverage the trade-off and design the proposed network while meeting some delay or SIR coverage constraints. 


\begin{table}
	\centering
	\caption{Trade-off Relationships By Changes of Parameters}\label{Table2}
	\begin{tabular}{|c|c|c|}
		\hline 
		parameters & short-term performance  & long-term performance \\ 
		\hline 
		increase $ \mu $ & decrease  & improve \\ 
		\hline 
			decrease $ \mu $ & improve & decrease \\ 
			\hline
		increase $\nu $  & decrease & improve \\ 
		\hline
				decrease $\nu $  & improve &  decrease\\ 
		\hline 
	\end{tabular} 
\end{table}

\subsection{Example: Network Optimization and Design}
 To elaborate on the processes of the network design and the optimization, we here consider a simple linear combination of the coverage probability and the area fraction. This is captured by the following aggregate utility 
\begin{equation}\label{utility}
	\mathcal{J}(\nu,\mu)=w_1 p_c(\tau) + w_2 \text{AF}(\bar{S}(\infty)),
\end{equation}
where $ p_c(\tau) $ is the coverage probability of the vehicle with SIR threshold $ \tau $, $ \text{AF}(\bar{S}(\infty)) $ is the limit of the area fraction of the cumulative coverage disks, $ w_1>0 $, and $ w_2>0 $. The coverage probability and the area fraction are both functions of $ \nu $ and $ \mu $. The above utility function is given by a linear combination of short-term and long-term metrics that behave differently with regards to each $ \nu $ and $ \mu $, respectively. See Table \ref{Table2}. 

\par A classical way of understanding $ \mathcal{J}(\nu,\mu) $ is to view it as a \emph{revenue}, or more simply the total utility of the netwo.rk, and to optimize it with respect to network parameters. For convenience, this paper assumes that the revenue or the utility would increase linearly with the values of the SIR coverage probability and the area fraction over time. The weights $ w_1,w_2 $ are positive numbers parameterizing the utility. As an example, if reliable communication is desirable at the expense of low latency, one can use a higher value $ w_1 $. Similarly, if one intends vehicles to sweep a wider area, a higher value is used for $ w_2 $. In general, depending on specific high-level design criteria and principles, different weights can be used. 
\par Given $ \mathcal{J} $, one can maximize the revenue or utility by jointly finding the best combination of coverage disks or inter-vehicle distance. Mathematically, for the given weights parameters $ w_1 $ and $ w_2 $, an unconstrained optimization problem is generally given by 
\begin{align}
\nu^\star,\mu^\star&=\argmax_{\nu,\mu > 0}\mathcal{J}(\nu,\mu)=\argmax_{\nu,\mu}\left(w_1 p_c(\tau) + w_2 \text{AF}(\overline{S}(\infty))\right).
\end{align}
In practice, on the other hand, such a formulation is maybe infeasible because the network parameters are often confined to certain ranges, due to various reasons such as implementation issues or radio regulations. Regardless of whether the optimization problem is constrained or not, one can analytically find the optima using existing methods, e.g., the Karush–Kuhn–Tucker condition\cite{boyd2004convex}. Yet, this paper only focuses on the numerical demonstration of the utility optimization.

\begin{figure}
	\centering
	\begin{minipage}{.5\textwidth}
		\centering
		\includegraphics[width=1\linewidth]{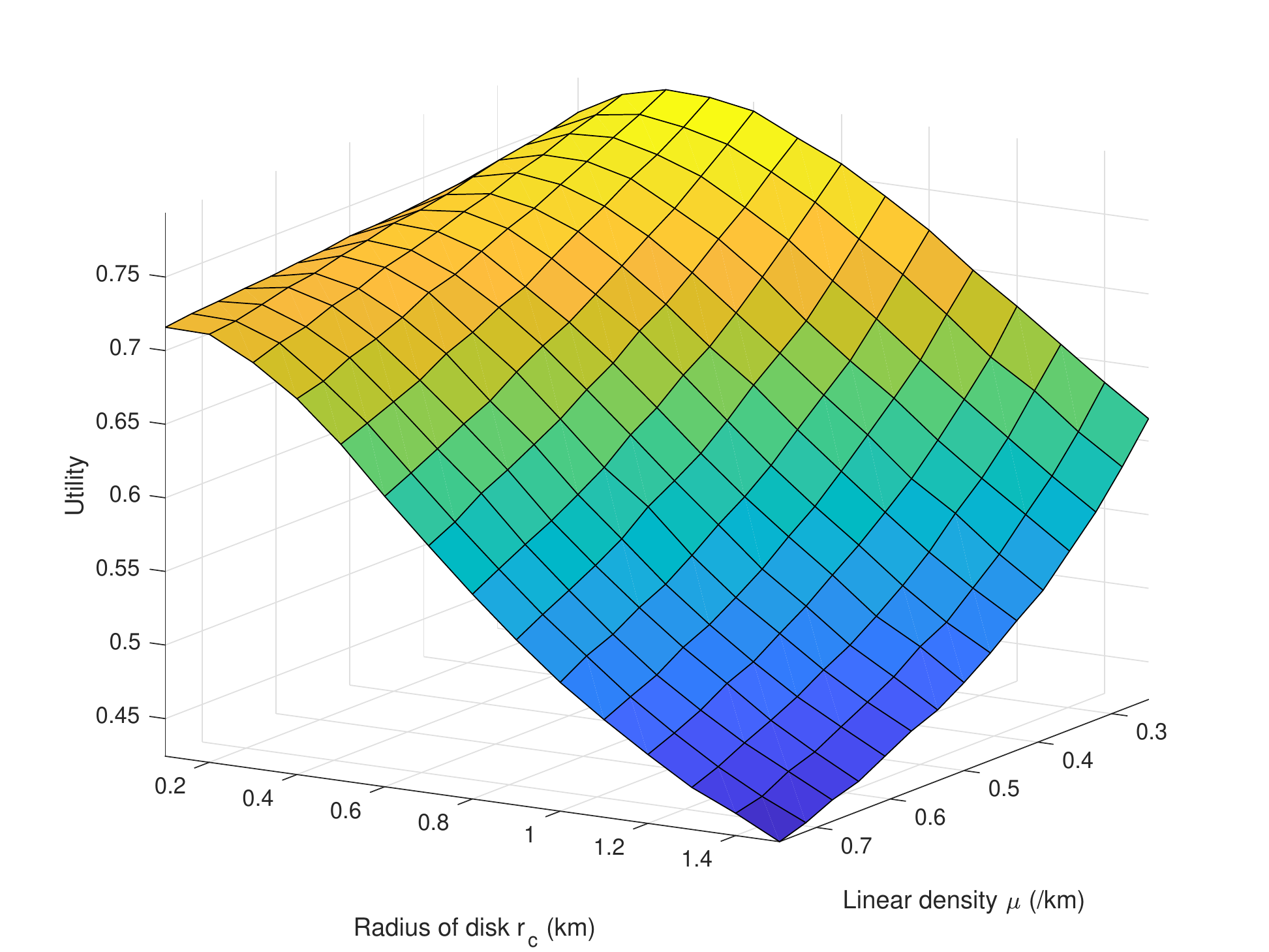}
	\end{minipage}%
	\begin{minipage}{.5\textwidth}
		\centering
		\includegraphics[width=1\linewidth]{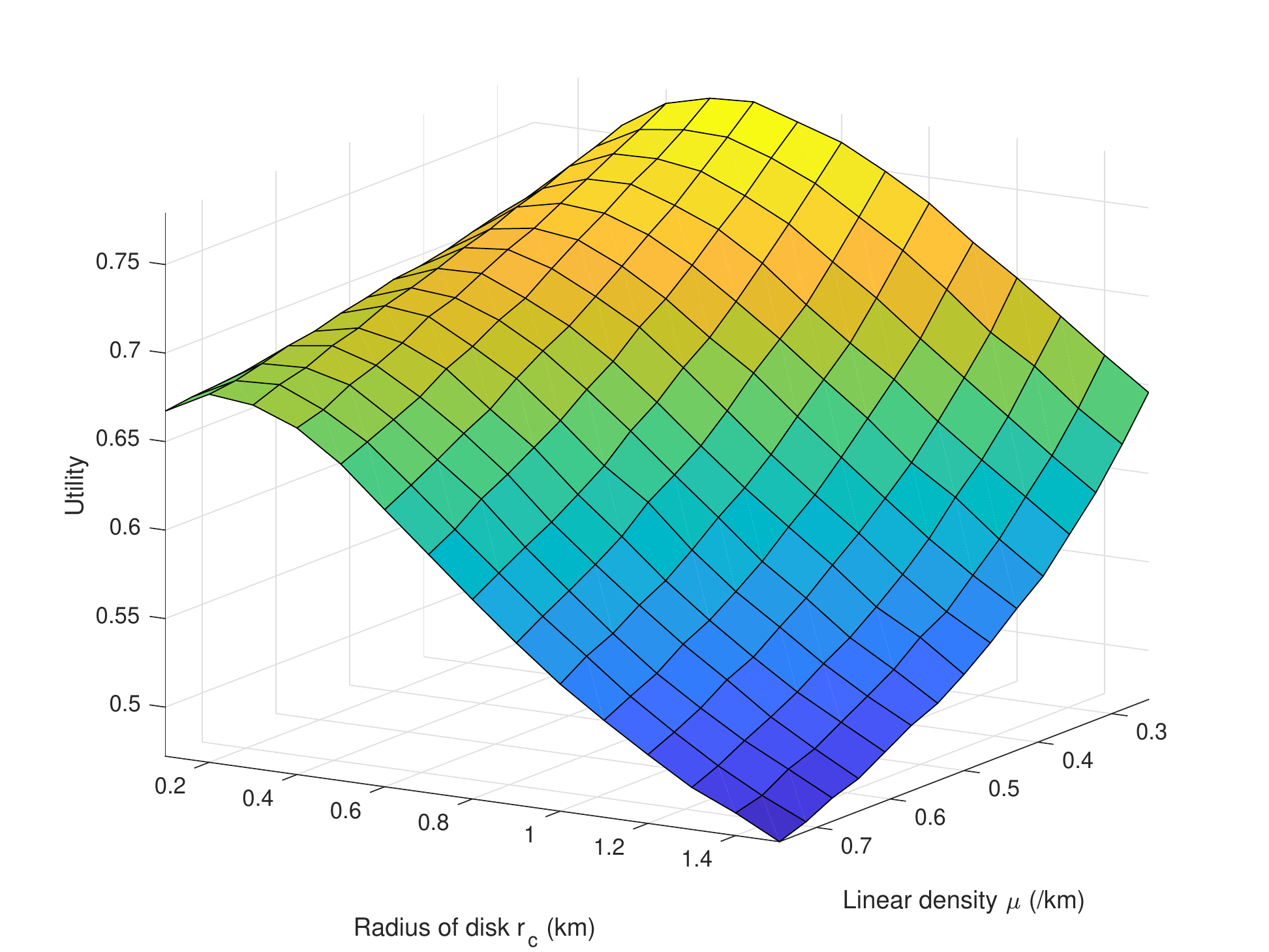}
	\end{minipage}
	\caption{Illustration of the utility. We consider $ w_1=0.7 $, $ w_2=0.3 $ for the left and $ w_1=0.64 $, $ w_2=0.36 $ for the right.}\label{fig:jointutility}
\end{figure}


\par Fig. \ref{fig:jointutility} illustrates the joint utility function of Eq \eqref{utility} with parameters $ \nu $ and $ \mu  $ on the $ x$-axis  and $ y$-axis, respectively. The coverage probability and area fraction of Eq. \eqref{utility} are directly obtained from Eqs. \eqref{eq:T1} and \eqref{25}, respectively. The left one considers $ w_1=0.7,w_2=0.3 $ and the right one considers $ w_1=0.64, w_2=0.36 $. Both cases illustrate the case where the proposed vehicular architecture is more focused on the reliability of links rather than the total area of coverage. In the left figure,  the joint utility is concave for given density $ \mu $ and thus, there exists a single value $ \nu^\star $ that maximize $ \mathcal{J}. $ On the other hand, for given radius $ \nu, $ the utility monotonically decreases with the coverage disk radii. Overall, for the given set $ \nu\in(0.1,1.5) $ and $ \mu\in(0.25,0.75), $ the utility achieves its maximum $ 0.8 $ for $ (\nu^\star,\mu^\star)=(0.4,0.39). $ In the right figure, the utility function is also concave for a given density $ \mu. $ Using the same principle, one obtains the best $ \nu^\star $ for a given  density $ \mu $ and the best pair is given by $ (\nu^\star,\mu^\star)=(0.5,0.39). $

\par
\begin{figure}
	\centering
	\includegraphics[width=0.7\linewidth]{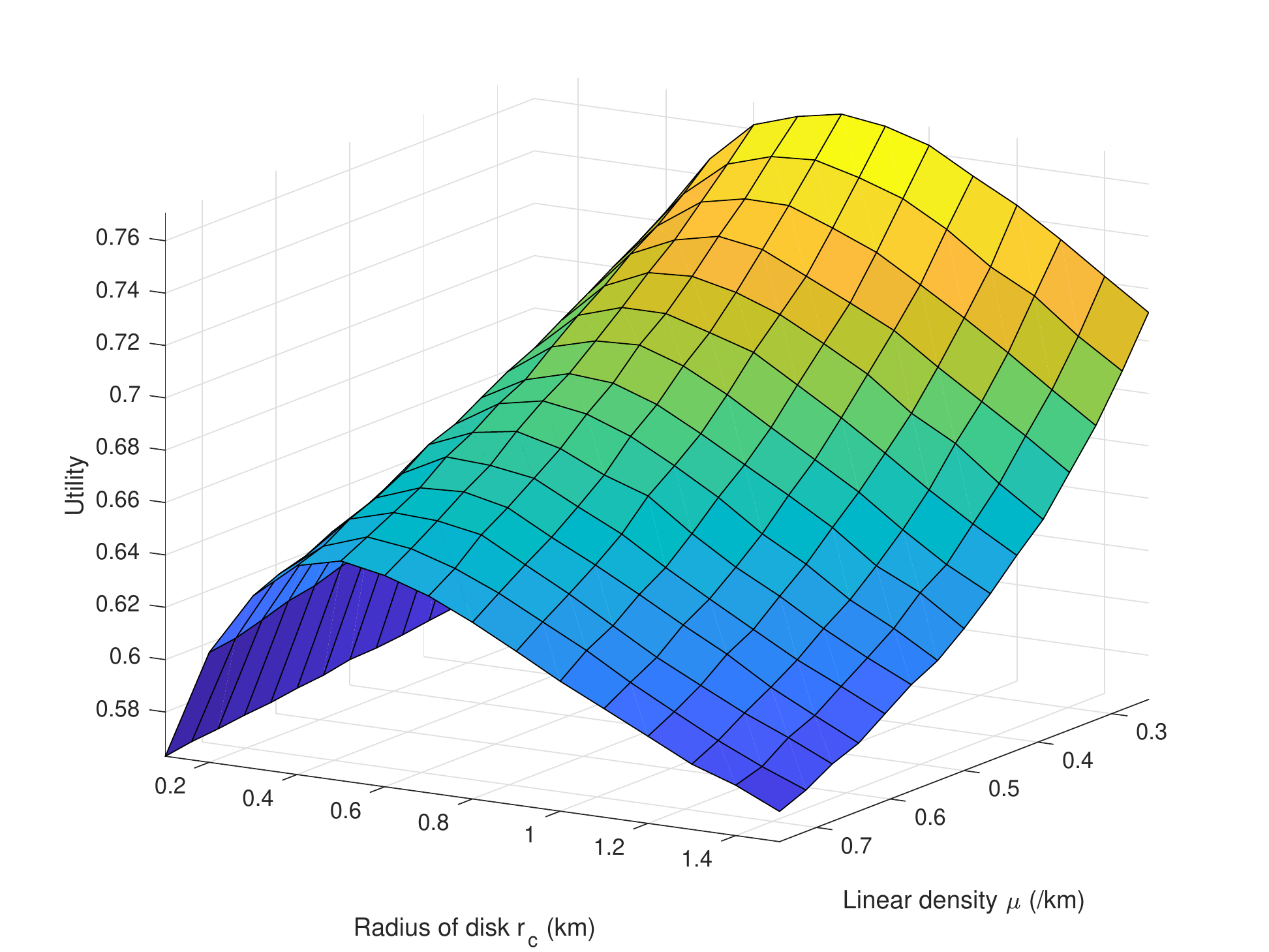}
	\caption{Illustration of the utility with weights $ w_1=w_2=0.5 $}
	\label{fig:jointutility3}
\end{figure}

 Fig. \ref{fig:jointutility3} illustrates the utility when the weights are equal, $ w_1=w_2=0.5. $ As in Fig. \ref{fig:jointutility}, the utility function is concave with respect to $ \nu $ for a given value of $ \mu. $ Consequently, the optimum pair $ \nu^\star,\mu^\star $ can be found similarly. 
\par In fact, any objective function, combining performance metrics with contrasting behaviors, e.g., short-term and long-term, can be used to provide a way to balance the various pros and cons of the proposed vehicular architecture. For instance, the optimization problem can be also written as
\begin{align}
\argmax_{\nu,\mu>0}\left\{  w_1 p_c(\tau) + w_2 \text{AF}(\bar{S}(\infty)) \right\} \text{ subject to } \bE[W|0\in\bar{S}(\infty)]<C.
\end{align}
Now the optimization problem is constrained by the delay, or equivalently the latency. Using Theorem \ref{T:4}, the optimal pair $(\nu^\star,\mu^\star) $ could be  found over the new domain 
\begin{equation*}
	Dom(\nu,\mu)=\left\{(\nu,\mu)\in\bR^2{}\left|\int_{0}^{\infty}\exp\left(-2\lambda_l\int_{0}^{\nu}1-e^{-2\mu \left(\sqrt{\nu^2-u^2}+vw\right)}\diff u\right)\diff w< C\right.\right\}.
\end{equation*}
The set can be numerically obtained by using the methods for integral equation e.g., Taylor expansion \cite{atkinson2009numerical}. Another way of capturing the utility with some latency constraint is to use negative utility for latency 
\begin{equation}
	\argmax_{\nu,\mu>0} \left\{w_1 p_c(\tau) + w_2 \text{AF}(\overline{S}(\infty)) -w_3\bE[W|0\in\bar{S}(\infty)]\right\},\label{243}
\end{equation}
where now the waiting time decrease the total utility as long as $ w_3>0 $. A detailed analysis of the network optimization including the latency aspect is left for future work. 
\subsection{Comparison: Typical Cellular Architecture}
Here, we assess the proposed network by contrasting its benefits and disadvantages to those of a general network, e.g., a network with static harvesters. In particular, we use uplink communications in the cellular network as a benchmark. The main purpose is to provide a qualitative comparison in order to give insights and to assess its potential as a large-scale connectivity architecture. 

The performance metrics we compare are the SINR coverage probability, the area spectral efficiency, the area fraction of the cumulative coverage disks (i.e., coverage region), and the network latency. The SINR coverage and the area spectral efficiency rate distribution capture the reliability and throughput of the communications present in the architecture. The area fraction corresponds to the area within which base stations provide the connectivity and, for this reason, is equivalent to the coverage region. 
\par 
To study their differences, the comparable parameters are made consistent for the two architectures; for instance, each hexagon cell has one uplink device at some time and the location of the latter is uniformly distributed inside the cell.  Fading distributions follow the same exponential distribution. No shadowing is considered. Notice that other aspects unrelated to the wireless performance and connectivity---e.g., the cost of establishing back-hauls for base stations, connecting vehicles, or even driving them---are out of the scope of this paper and thus they are not considered in this comparison. 
\par Table \ref{tableII} summarizes the performances of the two architectures. For the vehicular network, we use Theorems \ref{T:1} -- \ref{T:4} in Sections \ref{S:3} and \ref{S:4}. On the other hand, to evaluate the SINR distribution and rate distribution of the cellular network, we use \cite{novlan2013analytical}. Notice that the coverage region is almost equal to one because the hexagonal cellular network blankets the surface with a marginal cell edge region. Furthermore, we assess the latency of classical cellular networks as arbitrary small because, under any best-effort protocol, an uplink mobile device can be {scheduled} almost immediately. 
\par In the comparison with existing cellular architectures, the proposed network is shown to have both strengths and weaknesses. As Table \ref{tableII} shows, the cellular network has a more favorable coverage region and latency. On the other hand, given the variability of traffic level, devices might face significant delays due to fundamental limitation given by the geometry.  Nevertheless, the proposed network is shown to have better SINR coverage and area spectral efficiency. Because as long as the radius of the coverage disk is small enough, the proposed network is able to establish a short and reliable communication link.

\begin{table}
	\centering
	\caption{Qualitative Performance Comparison of Two Architectures}\label{tableII}
	\begin{tabular}{|c|c|c|}
		\hline 
		performance & proposed vehicular network  & cellular uplink network  \\ 
		\hline 
		SINR coverage probability & Theorem \ref{T:1} & \cite{novlan2013analytical} \\ 
		\hline 
		rate distribution & Theorem \ref{T:2} &  \cite{novlan2013analytical} \\ 
		\hline 
		coverage region & Theorem \ref{T:3} & $ \approxeq 1 $   \\ 
		\hline 
		Best latency & Theorem \ref{T:4} & $ \approxeq 0 $ \\ 
		\hline 
	\end{tabular} 
\end{table}

\section{Conclusion}\label{S:6}
In this paper, we propose a new network architecture where vehicles collect data from devices randomly distributed in space. The proposed system leverages the idea that moving vehicles can communicate, even if briefly, with a significant number of roadside devices and thus can provide wide connectivity. We analyze the proposed network in both the space and time domain. First, we analyze the SIR coverage probability and the area spectral efficiency based on a snapshot of the network geometry. Then, we investigate the evolution of the geometry to derive the area fraction of the coverage disks and the latency. We discuss various trade-off relationships and present the optimization of the utility function based on these performance metrics. We also compare the proposed network with the existing cellular architecture. 

\bibliographystyle{IEEEtran}
\bibliography{coxcox_ref}

\begin{thebibliography}{10}
\providecommand{\url}[1]{#1}
\csname url@samestyle\endcsname
\providecommand{\newblock}{\relax}
\providecommand{\bibinfo}[2]{#2}
\providecommand{\BIBentrySTDinterwordspacing}{\spaceskip=0pt\relax}
\providecommand{\BIBentryALTinterwordstretchfactor}{4}
\providecommand{\BIBentryALTinterwordspacing}{\spaceskip=\fontdimen2\font plus
\BIBentryALTinterwordstretchfactor\fontdimen3\font minus
  \fontdimen4\font\relax}
\providecommand{\BIBforeignlanguage}[2]{{%
\expandafter\ifx\csname l@#1\endcsname\relax
\typeout{** WARNING: IEEEtran.bst: No hyphenation pattern has been}%
\typeout{** loaded for the language `#1'. Using the pattern for}%
\typeout{** the default language instead.}%
\else
\language=\csname l@#1\endcsname
\fi
#2}}
\providecommand{\BIBdecl}{\relax}
\BIBdecl

\bibitem{ccs}
C.~Choi and F.~Baccelli, ``Modeling and optimization of direct communications
  from {IoT} devices to vehicles,'' in \emph{Proc. IEEE Globecom}, 2018.

\bibitem{forstall2013mobile}
S.~Forstall, G.~N. Christie, R.~E. Borchers, and K.~Tiene, ``Mobile device base
  station,'' Jun 2013, {U.S.} Patent 8,463,238.

\bibitem{saad2014vehicle}
E.~W. Saad, J.~L. Vian, M.~A. Vavrina, J.~A. Nisbett, and D.~C. Wunsch,
  ``Vehicle base station,'' Dec 2014, {U.S.} Patent no. 8,899,903.

\bibitem{talluri2018enhanced}
M.~Talluri, K.~Agarwal, R.~K. Mishra, and S.~Garg, ``Enhanced mobile base
  station,'' Mar. 2018, {U.S.} Patent no. 9,913,095.

\bibitem{Hull:2006:CDM:1182807.1182821}
B.~Hull, V.~Bychkovsky, Y.~ehang, K.~Chen, M.~Goraczko, A.~Miu, E.~Shih,
  H.~Balakrishnan, and S.~Madden, ``Cartel: A distributed mobile sensor
  computing system,'' in \emph{Proc. SenSys}, 2006.

\bibitem{jain2006exploiting}
S.~Jain, R.~C. Shah, W.~Brunette, G.~Borriello, and S.~Roy, ``Exploiting
  mobility for energy efficient data collection in wireless sensor networks,''
  \emph{Mobile Networks and Applications}, vol.~11, no.~3, pp. 327--339, 2006.

\bibitem{Xing:2008:RDA:1374618.1374650}
G.~Xing, T.~Wang, W.~Jia, and M.~Li, ``Rendezvous design algorithms for
  wireless sensor networks with a mobile base station,'' in \emph{Proc. ACM
  MobiHoc}, 2008, pp. 231--240.

\bibitem{Choi:2018:DLM:3209582.3209590}
C.~Choi, F.~Baccelli, and G.~de~Veciana, ``Densification leveraging mobility:
  An {IoT} architecture based on mesh networking and vehicles,'' in \emph{Proc.
  ACM MobiHoc}, 2018, pp. 71--80.

\bibitem{Veniam}
``Veniam: An {I}nternet of moving things,'' \url{http://veniam.com}, accessed:
  2018-11-11.

\bibitem{sugimoto2008prototype}
C.~Sugimoto, Y.~Nakamura, and T.~Hashimoto, ``Prototype of
  pedestrian-to-vehicle communication system for the prevention of pedestrian
  accidents using both {3G} wireless and {WLAN} communication,'' in \emph{Proc.
  IEEE ISWPC}, 2008, pp. 764--767.

\bibitem{anaya2014vehicle}
J.~J. Anaya, P.~Merdrignac, O.~Shagdar, F.~Nashashibi, and J.~E. Naranjo,
  ``Vehicle to pedestrian communications for protection of vulnerable road
  users,'' in \emph{Proc. IEEE Intell. Veh. Symp.}, 2014, pp. 1037--1042.

\bibitem{baccelli2010stochastic}
F.~Baccelli, B.~B{\l}aszczyszyn \emph{et~al.}, ``Stochastic geometry and
  wireless networks: Volume {I} theory,'' \emph{Foundations and
  Trends{\textregistered} in Networking}, vol.~3, no. 3--4, pp. 249--449, 2010.

\bibitem{chiu2013stochastic}
S.~N. Chiu, D.~Stoyan, W.~S. Kendall, and J.~Mecke, \emph{Stochastic geometry
  and its applications}.\hskip 1em plus 0.5em minus 0.4em\relax John Wiley \&
  Sons, 2013.

\bibitem{hartenstein2008tutorial}
H.~Hartenstein and L.~Laberteaux, ``A tutorial survey on vehicular ad hoc
  networks,'' \emph{IEEE Commun. Mag.}, vol.~46, no.~6, 2008.

\bibitem{huang2009spectrum}
K.~Huang, V.~K. Lau, and Y.~Chen, ``Spectrum sharing between cellular and
  mobile ad hoc networks: transmission-capacity trade-off,'' \emph{IEEE J. Sel.
  Areas Commun.}, vol.~27, no.~7, pp. 1256--1267, Sept. 2009.

\bibitem{doppler2009device}
K.~Doppler, M.~Rinne, C.~Wijting, C.~B. Ribeiro, and K.~Hugl,
  ``Device-to-device communication as an underlay to {LTE}-advanced networks,''
  \emph{IEEE Commun. Mag.}, vol.~47, no.~12, Dec 2009.

\bibitem{golrezaei2013femtocaching}
N.~Golrezaei, A.~F. Molisch, A.~G. Dimakis, and G.~Caire, ``Femtocaching and
  device-to-device collaboration: a new architecture for wireless video
  distribution,'' \emph{IEEE Commun. Mag.}, vol.~51, no.~4, pp. 142--149, Apr.
  2013.

\bibitem{feng2013device}
D.~Feng, L.~Lu, Y.~Yuan-Wu, G.~Y. Li, G.~Feng, and S.~Li, ``Device-to-device
  communications underlaying cellular networks,'' \emph{IEEE Trans. Commun.},
  vol.~61, no.~8, pp. 3541--3551, Aug. 2013.

\bibitem{andreev2014cellular}
S.~Andreev, A.~Pyattaev, K.~Johnsson, O.~Galinina, and Y.~Koucheryavy,
  ``Cellular traffic offloading onto network-assisted device-to-device
  connections,'' \emph{IEEE Commun. Mag.}, vol.~52, no.~4, pp. 20--31, Apr.
  2014.

\bibitem{weber2005transmission}
S.~P. Weber, X.~Yang, J.~G. Andrews, and G.~De~Veciana, ``Transmission capacity
  of wireless ad hoc networks with outage constraints,'' \emph{IEEE Trans. Inf.
  Theory}, vol.~51, no.~12, pp. 4091--4102, 2005.

\bibitem{baccelli2006aloha}
F.~Baccelli, B.~Blaszczyszyn, and P.~Muhlethaler, ``An aloha protocol for
  multihop mobile wireless networks,'' \emph{IEEE Trans. Inf. Theory}, vol.~52,
  no.~2, pp. 421--436, Feb. 2006.

\bibitem{andrews2010primer}
J.~G. Andrews, R.~K. Ganti, M.~Haenggi, N.~Jindal, and S.~Weber, ``A primer on
  spatial modeling and analysis in wireless networks,'' \emph{IEEE Commun.
  Mag.}, vol.~48, no.~11, Nov. 2010.

\bibitem{lin2014overview}
X.~Lin, J.~Andrews, A.~Ghosh, and R.~Ratasuk, ``An overview of {3GPP}
  device-to-device proximity services,'' \emph{IEEE Commun. Mag.}, vol.~52,
  no.~4, pp. 40--48, Apr. 2014.

\bibitem{haenggi2009stochastic}
M.~Haenggi, J.~G. Andrews, F.~Baccelli, O.~Dousse, and M.~Franceschetti,
  ``Stochastic geometry and random graphs for the analysis and design of
  wireless networks,'' \emph{IEEE J. Sel. Areas Commun.}, vol.~27, no.~7, pp.
  1029--1046, Sept. 2009.

\bibitem{baccelli2009stochastic}
F.~Baccelli, B.~Blaszczyszyn, and P.~Muhlethaler, ``Stochastic analysis of
  spatial and opportunistic aloha,'' \emph{IEEE J. Sel. Areas Commun.},
  vol.~27, no.~7, pp. 1105--1119, Sept. 2009.

\bibitem{ganti2009spatial}
R.~K. Ganti and M.~Haenggi, ``Spatial and temporal correlation of the
  interference in {ALOHA} ad hoc networks,'' \emph{IEEE Commun. Letters},
  vol.~13, no.~9, pp. 631--633, Sept. 2009.

\bibitem{moller2012lectures}
J.~Moller, \emph{Lectures on random Voronoi tessellations}.\hskip 1em plus
  0.5em minus 0.4em\relax Springer, 2012, vol.~87.

\bibitem{baccelli1997stochastic}
F.~Baccelli and S.~Zuyev, ``Stochastic geometry models of mobile communication
  networks,'' \emph{Frontiers in Queueing}, pp. 227--243, 1997.

\bibitem{morlot2012population}
F.~Morlot, ``A population model based on a {P}oisson line tessellation,'' in
  \emph{Proc. IEEE WiOpt}, 2012, pp. 337--342.

\bibitem{choi2017analytical}
C.~Choi and F.~Baccelli, ``An analytical framework for coverage in cellular
  networks leveraging vehicles,'' \emph{IEEE Trans. Commun.}, vol.~66, no.~10,
  pp. 4950--4964, Oct 2018.

\bibitem{chetlur2017coverage}
V.~V. Chetlur and H.~S. Dhillon, ``Coverage analysis of a vehicular network
  modeled as {C}ox process driven by {P}oisson line process,'' \emph{IEEE
  Trans. Wireless Commun.}, vol.~17, no.~7, pp. 4401--4416, July 2018.

\bibitem{molchanov2005theory}
I.~Molchanov, \emph{Theory of random sets}.\hskip 1em plus 0.5em minus
  0.4em\relax Springer, 2005, vol.~87, no.~2.

\bibitem{grossglauser2002mobility}
M.~Grossglauser and D.~N. Tse, ``Mobility increases the capacity of ad hoc
  wireless networks,'' \emph{IEEE/ACM Trans. Netw.}, vol.~10, no.~4, pp.
  477--486, Aug 2002.

\bibitem{zorzi2003geographic}
M.~Zorzi and R.~R. Rao, ``Geographic random forwarding ({GeRaF}) for ad hoc and
  sensor networks: energy and latency performance,'' \emph{IEEE Trans. Mobile
  Comput.}, vol.~2, no.~4, pp. 349--365, Oct 2003.

\bibitem{fall2008dtn}
K.~Fall and S.~Farrell, ``{DTN}: an architectural retrospective,'' \emph{IEEE
  J. Sel. Areas Commun.}, vol.~26, no.~5, pp. 828--836, June 2008.

\bibitem{skordylis2008delay}
A.~Skordylis and N.~Trigoni, ``Delay-bounded routing in vehicular ad-hoc
  networks,'' in \emph{Proc. ACM Mobihoc}, 2008, pp. 341--350.

\bibitem{pereira2012delay}
P.~R. Pereira, A.~Casaca, J.~J. Rodrigues, V.~N. Soares, J.~Triay, and
  C.~Cervello-Pastor, ``From delay-tolerant networks to vehicular
  delay-tolerant networks,'' \emph{IEEE Commun. Surv\&Tuts}, vol.~14, no.~4,
  pp. 1166--1182, Fourth 2012.

\bibitem{shah2003data}
R.~C. Shah, S.~Roy, S.~Jain, and W.~Brunette, ``Data mules: modeling a
  three-tier architecture for sparse sensor networks,'' in \emph{Proc. IEEE
  Int. Workshop on Sensor Netw. Protocols and Applicat., 2003.}, May 2003, pp.
  30--41.

\bibitem{spyropoulos2005spray}
T.~Spyropoulos, K.~Psounis, and C.~S. Raghavendra, ``Spray and wait: an
  efficient routing scheme for intermittently connected mobile networks,'' in
  \emph{Proc. ACM SIGCOMM workshop on Delay-tolerant networking}, 2005, pp.
  252--259.

\bibitem{zhao2005controlling}
W.~Zhao, M.~Ammar, and E.~Zegura, ``Controlling the mobility of multiple data
  transport ferries in a delay-tolerant network,'' in \emph{Proc. IEEE
  INFOCOM}, vol.~2, 2005, pp. 1407--1418.

\bibitem{choi2018poisson}
C.~Choi and F.~Baccelli, ``Poisson cox point processes for vehicular
  networks,'' \emph{IEEE Trans. Veh. Technol.}, vol.~67, no.~10, pp.
  10\,160--10\,165, Oct 2018.

\bibitem{daley2008introduction}
D.~Daley and D.~Vere-Jones, \emph{An introduction to the theory of point
  processes. volume {II}: general theory and structure. Probability and its
  applications}.\hskip 1em plus 0.5em minus 0.4em\relax Springer Berlin, 2008.

\bibitem{hamdi2010useful}
K.~A. Hamdi, ``A useful lemma for capacity analysis of fading interference
  channels,'' \emph{IEEE Trans. Commun.}, vol.~58, no.~2, pp. 411--416, Feb.
  2010.

\bibitem{gubbi2013internet}
J.~Gubbi, R.~Buyya, S.~Marusic, and M.~Palaniswami, ``Internet of things
  {(IoT)}: A vision, architectural elements, and future directions,''
  \emph{Future generation computer systems}, vol.~29, no.~7, pp. 1645--1660,
  2013.

\bibitem{boyd2004convex}
S.~Boyd and L.~Vandenberghe, \emph{Convex optimization}.\hskip 1em plus 0.5em
  minus 0.4em\relax Cambridge university press, 2004.

\bibitem{atkinson2009numerical}
K.~Atkinson and W.~Han, ``Numerical solution of fredholm integral equations of
  the second kind,'' in \emph{Theoretical Numerical Analysis}.\hskip 1em plus
  0.5em minus 0.4em\relax Springer, 2009, pp. 473--549.

\bibitem{novlan2013analytical}
T.~D. Novlan, H.~S. Dhillon, and J.~G. Andrews, ``Analytical modeling of uplink
  cellular networks,'' \emph{IEEE Trans. Wireless Commun.}, vol.~12, no.~6, pp.
  2669--2679, June 2013.

\end{thebibliography}
\appendices

\end{document}